\newif\ifsubmode
\submodefalse

\newif\ifprintfig
\printfigtrue

\ifsubmode
  \documentstyle[12pt,aasms4,epsf]{article}
  \received{}
  \accepted{}
  \journalid{}{}
  \articleid{}{}
\else
  \documentstyle[11pt,aaspp4,epsf]{article}
  \slugcomment{{\it submitted to The Astronomical Journal on April 19, 1998}}
\fi

\lefthead{van der Marel \& van den Bosch}
\righthead{A black hole in NGC 7052}

\newcommand{\etal}{{et al.~}}

\newcommand{\lta}{\lesssim}
\newcommand{\gta}{\gtrsim}

\newcommand{\kms}{\>{\rm km}\,{\rm s}^{-1}}
\newcommand{\pc}{\>{\rm pc}}
\newcommand{\Mpc}{\>{\rm Mpc}}
\newcommand{\Msun}{\>{\rm M_{\odot}}}
\newcommand{\Lsun}{\>{\rm L_{\odot}}}

\newcommand{\Mbh}{M_{\bullet}}
\newcommand{\Rt}{R_{\rm t}}
\newcommand{\Vc}{V_{\rm c}}

\newcommand{\halnii}{H$\alpha$+[NII]}

\newcommand{\BB}{$B$}
\newcommand{\VV}{$V$}

\newcommand{\II}{$I$}

\begin{document}

\title{Evidence for a $3 \times 10^8 \Msun$ black hole in NGC 7052\\
from HST observations of the nuclear gas disk\altaffilmark{1}}

\author{Roeland P.~van der Marel\altaffilmark{2,3,4}}
\affil{Space Telescope Science Institute, 3700 San Martin Drive, 
       Baltimore, MD 21218}

\author{Frank C.~van den Bosch\altaffilmark{4,5}}
\affil{Department of Astronomy, University of Washington, Box 351580, 
       Seattle, WA 98195-1580}


\altaffiltext{1}{Based on observations with the NASA/ESA Hubble Space 
       Telescope obtained at the Space Telescope Science Institute, which is 
       operated by the Association of Universities for Research in Astronomy, 
       Incorporated, under NASA contract NAS5-26555.}

\altaffiltext{2}{STScI Fellow.}

\altaffiltext{3}{Previously at the Institute for Advanced Study, 
       Olden Lane, Princeton, NJ 08540.}

\altaffiltext{4}{Hubble Fellow.}

\altaffiltext{5}{Previously at the Sterrewacht Leiden,
       Postbus 9513, 2300 RA, Leiden, The Netherlands.}


\ifsubmode\else
\clearpage\fi


\ifsubmode\else
\baselineskip=14pt
\fi


\begin{abstract}
We present a Hubble Space Telescope (HST) study of the nuclear region
of the E4 radio galaxy NGC 7052, which has a nuclear disk of dust and
gas. The Second Wide Field and Planetary Camera (WFPC2) was used to
obtain {\BB}, {\VV} and~{\II} broad-band images and an {\halnii}
narrow-band image. The images yield the stellar surface brightness
profile, the optical depth of the dust, and the flux distribution of
the ionized gas. The Faint Object Spectrograph (FOS) was used to
obtain {\halnii} spectra at six different positions along the major
axis, using a $0.26''$ diameter circular aperture. The emission lines
yield the rotation curve of the ionized gas and the radial profile of
its velocity dispersion. The observed rotation velocity at $r=0.2''$
from the nucleus is $V = 155 \pm 17 \kms$. The Gaussian dispersion of
the emission lines increases from $\sigma \approx 70\kms$ at $r=1''$,
to $\sigma \approx 400\kms$ on the nucleus.

To interpret the gas kinematics we construct axisymmetric models in
which the gas and dust reside in a disk in the equatorial plane of the
stellar body, and are viewed at an inclination of $70^{\circ}$. It is
assumed that the gas moves on circular orbits, with an intrinsic
velocity dispersion due to turbulence (or otherwise non-gravitational
motion). The latter is required to fit the observed increase in the
line widths towards the nucleus, and must reach a value in excess of
$500 \kms$ in the central $0.1''$. The circular velocity is calculated
from the combined gravitational potential of the stars and a possible
nuclear black hole. Models without a black hole predict a rotation
curve that is shallower than observed ($V_{\rm pred} = 92\kms$ at
$r=0.2''$), and are ruled out at $>99$\% confidence. Models with a
black hole of mass $\Mbh = 3.3^{+2.3}_{-1.3} \times 10^8 \Msun$
provide an acceptable fit. The best-fitting model with a black hole
adequately reproduces the observed emission line shapes on the
nucleus, which have a narrower peak and broader wings than a Gaussian.

NGC 7052 can be added to the list of active galaxies for which HST
spectra of a nuclear gas disk provide evidence for the presence of a
central black hole. The black hole masses inferred for M87, M84, NGC
6251, NGC 4261 and NGC 7052 span a range of a factor $10$, with NGC
7052 falling on the low end. By contrast, the luminosities of these
galaxies are identical to within $\sim\!25$\%. Any relation between
black hole mass and luminosity, as suggested by independent arguments,
must therefore have a scatter of at least a factor $10$.
\end{abstract}



\clearpage


\label{s:intro}
\section{Introduction}

Astronomers have been searching for direct evidence for the presence
of black holes (BHs) in galactic nuclei for more than two
decades. Initially, the only constraints on the central mass
distributions of galaxies were obtained from ground-based stellar
kinematical observations. More recently, the launch of the Hubble
Space Telescope (HST) and the subsequent refurbishment in 1993 have
provided an important increase in spatial resolution. Combined with
new techniques for data analysis and dynamical modeling this has
strengthened the stellar kinematical evidence for BHs in several
quiescent galaxies (e.g., Kormendy \etal 1996a,b; van der Marel \etal
1997a; Cretton \& van den Bosch 1998; Gebhardt \etal 1998). New tools
for the detection of BHs were also developed. HST observations of the
rotation velocities of nuclear disks of ionized gas provided accurate
BH mass determinations for several active galaxies (e.g., Harms \etal
1994; Ferrarese, Ford \& Jaffe 1996, 1998; Macchetto \etal 1997; Bower
\etal 1998), while for other galaxies BHs were detected through
VLBI observations of nuclear water maser sources (e.g., Miyoshi
\etal 1995). The case for a BH in our own galaxy improved drastically through
measurements of stellar proper motions exceeding $1000 \kms$ in the
central $0.1\pc$ (Genzel \etal 1997). There are now a total of 10---20
galaxies for which a nuclear dark mass, most likely a BH, has been
convincingly detected. The combined results for these galaxies are
summarized and reviewed in, e.g., Kormendy \& Richstone (1995), Ford
\etal (1998), Ho (1998), Richstone (1998), and van der Marel (1998). 
This sample is now large enough to study the BH mass distribution in
galaxies, which is further constrained by ground-based stellar
kinematical observations (Magorrian \etal 1998), HST photometry (van
der Marel 1998) and quasar evolution (e.g., Haehnelt \etal 1998). Our
understanding remains sketchy, but is consistent with a picture in
which a majority of galaxies has BHs, and in which the BH mass $\Mbh$
correlates with the luminosity or mass of the host spheroid.

In this paper we present and analyze HST data for the E4 galaxy NGC
7052. This galaxy is a radio source with a core and jet, but no lobes
(Morganti \etal 1987). Ground-based optical images show a nuclear dust
disk aligned with the major axis of the galaxy (Nieto \etal 1990). The
physical properties of this disk were discussed by de Juan, Colina \&
Golombek (1996). In a previous paper (van den Bosch \& van der Marel
1995; hereafter Paper~I) we presented ground-based narrow-band imaging
and long-slit spectroscopy obtained with the 4.2m William Herschel
Telescope (WHT). These observations showed that there is also a
rotating nuclear disk of ionized gas in NGC 7052.  The gas has a steep
central rotation curve, rising to nearly $300 \kms$ at $1''$ from the
center. However, the spatial resolution of the spectra was
insufficient to convincingly detect a BH, due in part to the
relatively large distance of NGC 7052 ($58.7 \Mpc$; i.e., $1'' = 284.6
\pc$). The velocity dispersion of the gas was found to increase from
$\sigma \approx 70\kms$ at $1''$ from the center to $\sigma \approx
200\kms$ on the nucleus. We showed that this cannot be the sole result
of rotational broadening, which would have predicted double-peaked
line profile shapes that are not observed.  Instead, the observed
central increase in the line width must be at least partly
intrinsic. The ground-based kinematics yield an upper limit of $\sim
10^9 \Msun$ on the mass of any possible BH.

To improve the constraints on the presence of a central BH we obtained
broad- and narrow-band images of NGC 7052 with the Second Wide Field
and Planetary Camera (WFPC2) and spectroscopy with the Faint Object
Spectrograph (FOS), both in the context of HST project GO-5848. We
discuss the imaging and photometric analysis in Section~\ref{s:WF},
and the spectroscopy and kinematical analysis in
Section~\ref{s:FOS}. In Section~\ref{s:dyn} we construct dynamical
models to interpret the results. With the high spatial resolution of
these data we are able to better constrain the nuclear mass
distribution, and we find that NGC 7052 has a BH with mass $\Mbh =
3.3^{+2.3}_{-1.3} \times 10^8 \Msun$. We summarize and discuss our
findings in Section~\ref{s:disc}. Some observational details are
presented in an appendix.

We adopt $H_0 = 80 \kms \Mpc^{-1}$ throughout this paper. This does
not directly influence the data-model comparison for any of our
models, but does set the length, mass and luminosity scales of the
models in physical units. Specifically, distances, lengths and masses
scale as $H_0^{-1}$, while mass-to-light ratios scale as $H_0$.

\section{WFPC2 observations and photometric analysis}
\label{s:WF}

\subsection{Setup and data reduction} 
\label{ss:WFobsdesc}

We used the WFPC2 to obtain both broad- and narrow-band images of NGC
7052. The WFPC2 is described in detail by, e.g., Biretta \etal
(1996). The observations took five spacecraft orbits in one `visit'.
Telescope tracking was done in `fine lock', with a RMS telescope
jitter of $\sim\! 3$ milli-arcsec (mas). The observing log is
presented in Table~\ref{t:images}. We obtained broad-band images with
the filters F450W, F547M and F814W, corresponding roughly to Johnson
{\BB}, {\VV} and~{\II}. We also obtained narrow-band images with the
`linear ramp filter' (LRF), for which the central wavelength varies as
function of position. This filter has a transmission curve with a FWHM
of $\sim\!1.3$\% of the central wavelength. By placing the target at a
known position on the detector, we obtained images of the nuclear
region around $6675${\AA} (`on-band'; covering {\halnii}) and
$6480${\AA} (`off-band').  The WFPC2 has four chips, each with $800
\times 800$ pixels. The {\BB} and~{\II} images were taken with the galaxy
centered on the PC chip, yielding a pixel size of $0.046''$. The {\VV}
and LRF images were taken with the galaxy centered on one of the WF
chips, yielding a pixel size of $0.100''$.

\placetable{t:images}

The images were calibrated by the HST calibration `pipeline'
maintained by the Space Telescope Science Institute (STScI). The
standard reduction steps include bias subtraction, dark current
subtraction and flat-fielding, as described in detail by Holtzman
\etal (1995a). The pipeline does not flat-field LRF images, so we manually  
flat-fielded those using the flat-field for a broad-band filter with a
comparable central wavelength (F675W). Three different exposures were
taken with each filter, each with the target shifted by a small
integer number of pixels. The different exposures for each filter were
aligned, and then combined with simultaneous removal of cosmic rays
and chip defects. This yields one final image per filter. The final
images were corrected for the geometric distortion of the WFPC2, and
constant backgrounds were subtracted, as measured at those regions of
the detector where the galactic contribution is negligible.

To obtain an image of the {\halnii} emission one must subtract an
off-band image from the on-band image. One option is to use the {\VV}
or {\II}-band image (or an appropriate combination) as off-band
image. These images have the advantage of high signal-to-noise ratio
($S/N$). However, the central wavelengths differ significantly from
that of {\halnii}, so the dust absorption in these images may not be
the same as in the on-band image. Such differential dust absorption
may be incorrectly interpreted as gas emission. To minimize
differential dust absorption, we instead used the $6480${\AA} LRF
image as off-band image. This off-band image was aligned with the
on-band image, scaled to fit the on-band image at radii where the
ionized gas flux is negligible, and subtracted from the on-band image.

The count-rates in the broad-band F450W, F547M and F814W images were
calibrated to magnitudes in the Johnson {\BB}, {\VV} and~{\II} bands,
respectively, as described in Holtzman \etal (1995b). The count-rates
in the LRF images were calibrated to erg cm$^{-2}$ s$^{-1}$ using
calculations with the SYNPHOT package in IRAF. These calculations used
the LRF transmission curve parameterizations given in Biretta \etal
(1996) and the spectral energy distribution around {\halnii} measured
in Paper I. The flux calibration of the {\halnii} image was verified
by comparing it to the results of our HST/FOS spectroscopy.

\subsection{Morphology}
\label{ss:WFmorph}

Figure~\ref{f:images} shows the reduced images, with from top-left to
bottom-right, the {\II}-band image, the {\BB}-band image, the $B-I$
image and the emission line image of {\halnii}. The most prominent
feature in the broad-band images is the absorption by the dust disk.
The morphology of the starlight in the central region resembles that
of a cone emanating from the nucleus. Since no such structure is seen
in the emission line image, we exclude the possibility that the cone
is related to an ionization cone. Rather, the cone may indicate that
the dust disk has a non-zero opening angle (such that its thickness
increases with distance). However, without detailed modeling it is not
possible to rule out alternative morphologies for the dust disk, such
as a constant scale-height or ring morphology. The $B-I$ image shows
the dust absorption more directly. The galaxy is reddest at the front
edge of the disk ($B-I \approx 3.0$), because this is where the
fraction of the starlight along the line of sight that is `eclipsed'
by the disk is largest. The dust lane is also visible in the {\halnii}
image, but less prominently than in the broad-band images.

\placefigure{f:images}

We have fitted by eye the outline of the dust disk in the $B-I$ image
and the outline of the gas disk in the {\halnii} image. Both are well
approximated by an ellipse, although the ionized gas has some
additional filamentary extensions to the north and west. The fitted
ellipses have a semi-major by semi-minor axis size of $1.94'' \times
0.67''$ for the dust disk, and $2.08'' \times 0.72''$ for the gas
disk. The disks have very similar size, probably indicating a common
origin. The inferred axial ratio is $q_{\rm disk} = 0.35 \pm 0.03$ for
both components, where the error was estimated by eye. If we make the
simplifying assumptions that both the dust and gas disks are circular,
have negligible thickness, and reside in the equatorial plane, then
the implied inclination is $i = \arccos q_{\rm disk} = (70 \pm
2)^{\circ}$. This is in excellent agreement with the results of the
models presented in paper I, which indicated $i = (70 \pm 5)^{\circ}$.

The centroid of the stellar isophotes at large radii coincides with
the observed peak surface brightness in the broad-band images to
within the errors of $\sim\!0.02''$. This, combined with the
point-like appearance of the `nucleus' in the broad-band images,
suggests that we may be seeing the center of the stellar distribution
through the dust. In contrast with the case for NGC 4261 (Ferrarese,
Ford \& Jaffe 1996), we have found no evidence (to within the errors
of $\sim\!0.04''$) that either the dust disk or the gas disk is offset
from the galaxy center defined by the starlight. The principal axes of
the stellar, gaseous and dust components are also well aligned. None
of the components is aligned with the position angle of the radio jet
in NGC 7052 (Morganti \etal 1987), as indicated in
Figure~\ref{f:images}. This has been found to be the case in more
galaxies with nuclear gas disks (de Juan, Colina \& Golombek 1996),
although in other galaxies the alignment with the radio axis has been
found to be quite good (e.g., NGC 4261 and M87; Ford \etal 1994;
Ferrarese, Ford \& Jaffe 1996).

The $B-I$ image shows two regions near the nucleus, labeled A and~B,
that are bluer ($B-I = 2.0$---$2.5$) than most of the dust obscured
region (which has $B-I = 2.6$---$3.0$) and also than the main body of
the galaxy. The blue color of these regions may be due to local star
formation or to a contribution of [OIII]5007 emission to the
{\BB}-band image (although no enhanced {\halnii} emission is seen at
these positions). The region~B is approximately aligned with the radio
jet of NGC 7052, and could also be non-thermal optical emission from a
jet knot. However, even the best available radio image of the jet
(Morganti \etal 1987) has a resolution of more than $1$ arcsec, so it
is unknown whether there is a radio knot at the position of the
region~B. Our ignorance of the exact nature of the regions~A and~B has
no impact on our study of the nuclear mass distribution of NGC 7052.

\subsection{The stellar luminosity density}
\label{ss:WFduststars}
 
The stellar surface brightness outside the dust distribution is best
analyzed by fitting ellipses to the isophotes. We have done this with
the software in the IRAF package STSDAS, which is based on the
algorithm of Jedrzejewski (1987). Figure~\ref{f:ellipsefit} shows the
{\II}-band surface brightness, the $B-V$ and $V-I$ colors, the
ellipticity ($\epsilon$), the major axis position angle ($\theta$) and
the amplitudes of the third and fourth order Fourier coefficients of
the isophotes (describing deviations from pure ellipses), all as
functions of radius along the major axis. Results are not shown for $r
\lesssim 2.0''$, where dust absorption dominates the observed
morphology.

\placefigure{f:ellipsefit}

The isophotal structure of NGC~7052 outside the dust disk is very
regular. The third and fourth order Fourier coefficients are
consistent with zero, indicating that the isophotes are accurately
elliptical. No significant isophote twist is present. There is only a
moderate increase in ellipticity, running from $\sim 0.3$ just outside
the dust disk, to $\sim 0.4$ at the edge of the HST images ($r \sim
15''$).

To study the stellar surface brightness distribution at small radii,
we consider the observed minor axis profile on the side of the galaxy
where the dust absorption is smallest (i.e., upward in
Figure~\ref{f:images}). Figure~\ref{f:minorprof} shows the {\II}-band
profile averaged over a $0.14''$ wide strip. The dust disk extends
$\sim 0.67''$ along the minor axis. The observed profile outside this
radius reflects the true stellar brightness profile, but at smaller
radii the brightness has been decreased by dust absorption.

\placefigure{f:minorprof}

To model the stellar surface brightness we adopt a parameterization for
the three-dimensional stellar luminosity density $j$. We assume that
$j$ is oblate axisymmetric, that the isoluminosity spheroids have
constant flattening $q$ as a function of radius, and that $j$ can be
parameterized as
\begin{equation}
  j(R,z) = j_0 (m/a)^{-\alpha} [1+(m/a)^2]^{-\beta} , \quad
  m^2 \equiv R^2 + z^2 q^{-2} .
\label{lumdendef}
\end{equation}
Here $(R,z)$ are the usual cylindrical coordinates, and $\alpha$,
$\beta$, $a$ and $j_0$ are free parameters. When viewed at inclination
angle $i$, the projected intensity contours are ellipses with axial
ratio $q'$, with $q'^2 \equiv \cos^2 i + q^2 \sin^2 i$. The projected
intensity for the luminosity density $j$ must be calculated
numerically. In the following we adopt $i=70^{\circ}$, based on the
observed shape of the dust and gas disks, and $q'=0.7$, as appropriate
for the region just outside the dust disk
(cf.~Figure~\ref{f:ellipsefit}). We neglect the slight increase in the
observed ellipticity with radius.

The minor axis {\II}-band profile does not strongly constrain the
central cusp slope $\alpha$, because of the dust absorption at small
radii. We therefore consider a one-parameter family of models with
different $\alpha$. For any fixed $\alpha$, the remaining model
parameters are determined by $\chi^2$ minimization to best fit the
minor axis {\II}-band profile outside the region influenced by dust
(between $0.67''$ and $4.0''$). In the data-model comparison we use
the projected surface brightness profile predicted by
equation~(\ref{lumdendef}), convolved with the WFPC2
point-spread-function (PSF) and pixel size. The PSF convolution is
important because the PSF wings extend to several arcseconds, despite
the fact that the FWHM is only $0.1''$. Dotted curves in the top panel
of Figure~\ref{f:minorprof} show the predictions for $\alpha=0.1$,
$0.2$, $0.3, \ldots, 1.3$. The bottom panel shows the residuals for
each fit, which must be attributed to dust absorption.

Models with larger values of $\alpha$ have more dust absorption, and
therefore predict more reddening of the starlight. So to further
constrain the models, and in particular $\alpha$, we consider the
observed $B-I$ color. Figure~\ref{f:minorcolor} shows $B-I$ as
function of minor axis distance, again averaged over a $0.14''$ wide
strip centered on the nucleus. Outside the central arcsec there is a
smooth color gradient, seen also in Figure~\ref{f:ellipsefit}. The
colors and color gradients of NGC 7052 have values typical of
elliptical galaxies. Color gradients such as seen in NGC 7052 are
believed to be due to stellar population gradients (e.g., Kormendy \&
Djorgovski 1989). Extrapolation of the outer $B-I$ gradient in NGC
7052 into the center emphasizes the additional reddening due to dust
absorption. The $B-I$ is largest at negative radii (downward in
Figure~\ref{f:images}), which are at the front side of the dust disk.
The reddening at positive radii is more modest. The nucleus itself is
slightly bluer than its surroundings (see also Figure~\ref{f:images}).
This may be an artifact due to differences in the PSFs for the {\BB}
and the {\II}-band images, but it may also indicate a contribution of
a nuclear non-thermal continuum or the presence of a central hole in
the dust distribution.

\placefigure{f:minorcolor}

To model the observed reddening we assume that the dust resides in the
equatorial plane. Let the $z$-axis be oriented such that the observer
sees the region $z>0$ in front of the dust disk, and let the $w$-axis
be along the line-of-sight. The {\II}-band surface brightnesses due to
the stars in front of and behind the dust are, respectively,
\begin{equation}
  I_{+} = \int_{z>0} j_I \> {\rm d}w , \qquad
  I_{-} = \int_{z<0} j_I \> {\rm d}w .
\label{projections}
\end{equation}
Here $j_I$ is the intrinsic {\II}-band luminosity density, which we
assume to be of the functional form given in
equation~(\ref{lumdendef}). The observed {\II}-band flux in pixel $i$
is
\begin{equation}
  {\widetilde I_{i}} = {\widetilde I_{+,i}} + {\widetilde I_{-,i}} 
  \exp[{-{\widehat \tau_{I,i}}}] ,
\label{obslumdensI}
\end{equation}
where the tildes denote convolution with the instrumental PSF and
pixel size. The quantity $\widehat \tau$ is the `effective' optical
depth that the light in pixel $i$ has encountered. Similarly, the
observed {\BB}-band flux in pixel $i$ is
\begin{equation}
  {\widetilde B_{i}} = {\widetilde B_{+,i}} + {\widetilde B_{-,i}} 
  \exp[{-{\widehat \tau_{B,i}}}] .
\label{predlumdensB}
\end{equation}

Given the {\II}-band image and a model for $j_I$, it is possible to
make predictions for the {\BB}-band image that can be compared to the
data. For a given $j_I$, we calculate ${\widetilde I_{+}}$ and
${\widetilde I_{-}}$ from equation~(\ref{projections}). We then assume
the stellar population gradient in $B-I$ to be given by the linear fit
in Figure~\ref{f:minorcolor}. Combined with ${\widetilde I_{+}}$ and
${\widetilde I_{-}}$ this yields predictions for ${\widetilde B_{+}}$
and ${\widetilde B_{-}}$ (neglecting the small differences in the PSFs
for the {\BB} and the {\II}-band images). The observed {\II}-band flux
yields ${\widehat \tau_{I}}$ from equation~(\ref{obslumdensI}). For
galactic dust, $\tau_B/\tau_I = 2.3347$ (Cardelli, Clayton \& Mathis
1989). Assuming that this ratio also applies to the effective optical
depths in NGC 7052, one obtains predictions for ${\widehat
\tau_{B}}$. Combined with ${\widetilde B_{+}}$ and ${\widetilde
B_{-}}$, equation~(\ref{predlumdensB}) then yields the predicted
{\BB}-band flux.

This scheme yields predictions $B_{\rm pred}$ for the {\BB}-band
fluxes for any assumed value of the central cusp slope $\alpha$. These
predictions can be compared to the observed fluxes $B_{\rm obs}$ using
the quantity
\begin{equation}
  \chi^2_{\alpha} = \sum_i \Biggl( {B_{{\rm pred},i} - B_{{\rm obs},i} \over
  \Delta B_{{\rm obs},i}} \Biggr)^2 .
\label{chi2reddening}
\end{equation}
In the summation we use pixels for which the projected radius $[x^2 +
y^2]^{1/2}$ exceeds $0.2''$ (to exclude pixels that may be influenced
by PSF differences between the {\BB} and {\II}-bands), and for which
the radius $[x^2 + (y/\cos i)^2]^{1/2}$ in the equatorial plane is
less than $2.0''$ (to exclude regions where there is no dust). Also,
the comparison is limited to only one side of the dust disk (the right
half in Figure~\ref{f:images}; this excludes the regions A and B which
may be contaminated by [OIII] emission). This yields a total of 995
pixels. The resulting $\chi^2_{\alpha}$ is shown as function of
$\alpha$ in the left panel of Figure~\ref{f:reddening}. The minimum
$\chi^2_{\alpha}$ is reached for $\alpha = 0.5$. Formal $\Delta
\chi^2$-statistics yield a very stringent constraint on the cusp
steepness ($0.45 \lta \alpha \lta 0.65$ at $99.73$\%
confidence). However, we are reluctant to accept this constraint at
face value; many assumptions underly our approach, and even the
minimum $\chi^2_{\alpha}$ is large for the number of degrees of
freedom. Instead, we prefer to be conservative, and do not wish to
rule out any model with $0.0 \leq
\alpha \lta 1.0$ solely on the basis of $\chi^2_{\alpha}$.

\placefigure{f:reddening}

It is unphysical to find $B_{\rm obs} < {\widetilde B_{+}}$ in a given
pixel; this is not even possible for infinite optical
depth. Nonetheless, one may expect a small fraction of the pixels to
have $B_{\rm obs} < {\widetilde B_{+}}$, due to the effects of Poisson
noise. The right panel in Figure~\ref{f:reddening} shows the fraction
of all pixels that has $B_{\rm obs} < {\widetilde B_{+}}$. This
fraction grows rapidly with increasing $\alpha$, and becomes
unacceptable for $\alpha \gta 0.8$. We therefore conclude overall from
our analysis of the reddening that $\alpha \lta 0.8$, with a best fit
for $\alpha = 0.5$.

Many groups have used the HST to study the surface brightness profiles
of elliptical galaxies (e.g., Crane \etal 1993; Jaffe \etal 1994;
Carollo \etal 1997; Faber \etal 1997). Faber \etal discuss the results
for 61 galaxies, using the so-called `NUKER law' parameterization for
the projected surface brightness profiles (Byun \etal 1996). To enable
a comparison with their results we have fitted NUKER laws to the
projected intensity profiles of our models over the region $0.1'' < r
< 10''$. Figure~\ref{f:centralslope} shows the central projected
intensity slope $\gamma$ of the NUKER law profile, as function of the
central luminosity density slope $\alpha$ of our models. At
asymptotically small radii one would expect $\gamma =
\max(0,\alpha-1)$, but we find $\gamma > \max(0,\alpha-1)$
at the radii observable with HST (consistent with the findings of
Gebhardt \etal 1996). The V-band absolute magnitude of NGC 7052 is
$M_V = -22.15$. Faber \etal find that elliptical galaxies can
be classified in two groups according to their surface brightness
profiles, and that all galaxies in their sample with $M_V < -22$ can
be classified as so-called `core galaxies'. The maximum $\gamma$ for
the 25 core galaxies in their sample is $0.25$. It is therefore
reasonable to assume that NGC 7052 also is a core galaxy with $\gamma
\lesssim 0.25$, which implies that $\alpha \lta 1.0$, 
(cf.~Figure~\ref{f:centralslope}). The NUKER law slope for $\alpha =
0.5$ is $\gamma = 0.09$, which is equal to the average for all the
core galaxies in the Faber \etal sample. The constraints on $\alpha$
in NGC 7052 inferred from our WFPC2 images are therefore consistent
with our understanding of elliptical galaxies as a class.

\placefigure{f:centralslope}

In the following we adopt the model with $\alpha = 0.5$ as our
`standard model'. The remaining parameters of this model are $\beta =
0.96$, $a = 1.10''$ and $j_0 = 14.8 \Lsun \pc^{-3}$ ({\II}-band),
which best fit the surface brightness profile at large radii. We
restrict much of the following discussion to our standard model, but
return to models with other values of $\alpha$ in
Section~\ref{ss:nonstandard}. In particular, we show there that
changes in the value of $\alpha$ cannot remove the need for a BH in
NGC 7052.

\subsection{The ionized gas disk}
\label{ss:WFgas}

Figure~\ref{f:gasfit} shows the {\halnii} flux along the major axis,
the intermediate axis and the minor axis of the gas disk,
respectively. The left panel shows the PSF profile for comparison. The
flux distribution is sharply peaked towards the nucleus. However, the
peak is broader than the PSF, so there is no evidence for an
unresolved nuclear point source.

\placefigure{f:gasfit}

For use in Section~\ref{s:dyn} we require a model for the flux
distribution. Given the lack of information on the distribution of the
ionized gas along the line of sight, we assume that it resides in an
infinitesimally thin circular disk in the equatorial plane of the
galaxy, and we parameterize the {\halnii} flux $F(R)$ as a truncated
double exponential:
\begin{equation}
  F(R) = \cases{
     F_1 \exp(-R/R_1) + F_2 \exp(-R/R_2) , & $(R \leq {\bar R})$ ,\cr
     0                                   , & $(R > {\bar R})$ .\cr}
\label{fluxparam}
\end{equation}
Observational information on the {\halnii} flux is available not only
from the WFPC2 image, but also from the spectra obtained with the FOS
(Section~\ref{s:FOS}) and the WHT (Paper~I). The best-fit parameters
in equation~(\ref{fluxparam}) were determined using $\chi^2$
minimization to optimize the fit to all observational constraints on
the emission line flux simultaneously. For the WFPC2 data we included
only the one-dimensional profiles shown in Figure~\ref{f:gasfit},
rather than the entire two-dimensional image. Convolutions with the
PSF and pixel and/or aperture size for each setup were properly taken
into account.

The predictions of the best-fit model are shown as solid curves in
Figure~\ref{f:gasfit}. The overall fit to the data is satisfactory,
and our simple model is sufficient for the dynamical interpretation of
the gas kinematics in Section~\ref{s:dyn}.  Nonetheless, two remarks
must be made.  First, the model is axisymmetric, so the features in
the data that are not symmetric with respect to the nucleus (e.g., the
bump at $\sim -1.0''$ along the major axis) cannot be
reproduced. Second, there is some inconsistency between the central
fluxes in the different data sets. The predicted and observed fluxes
for the spectra are shown in Figure~\ref{f:modelfits}. The HST/FOS
fluxes suggest a more sharply peaked flux distribution than the WFPC2
and WHT data, and our best fit model is a compromise between all the
data. Our model therefore overpredicts the central flux measured in
the WFPC2 data. The parameters of the best-fit model are: $R_1 =
0.032''$, $R_2 = 0.84''$, ${\bar R} = 2.02''$, and $F_1/F_2 =
140$. The value of $F_1$ is not important for our models, but only
sets the absolute normalization of the flux. The absolute calibration
of our data is most accurate for the HST/FOS spectra, from which we
derive $F_1 = 3.2 \times 10^{-12}$ erg cm$^{-2}$ s$^{-1}$
arcsec$^{-2}$.


\section{FOS observations and kinematical analysis}
\label{s:FOS}

\subsection{Setup and data reduction} 
\label{ss:FOSobsetup}

We obtained spectra of NGC 7052 in visits on September 7, 1995 and
August 18, 1996, using the red side detector of the HST/FOS. The
instrument is described in detail in Keyes \etal (1995). The COSTAR
optics corrected the spherical aberration of the HST primary
mirror. The G570H grating was used in `quarter-stepping' mode,
yielding spectra with 2064 pixels covering the wavelength range from
4569 {\AA} to 6818 {\AA}. All spectra were obtained with a $0.26''$
diameter circular aperture (the FOS {\tt 0.3} aperture).

Each visit consisted of seven spacecraft orbits, with each orbit
having $54$ minutes of target visibility time followed by $42$ minutes
of Earth occultation. The first two orbits in each visit were used for
target acquisition. Subsequent orbits were used to obtain spectra at
various positions near the nucleus. Periods of Earth occultation were
used to obtain wavelength calibration spectra of the internal arc
lamp. In part of the last orbit of each visit the FOS was used in a
special mode to obtain an image of the central part of NGC 7052, to
verify the telescope pointing.

We performed the target acquisition in each visit through a `peak-up'
on a nearby star (magnitude $V=17.4$; coordinates in the HST Guide
Star system: RA $=$ 21h 18m 32.46s, $\delta = 26^{\circ} \, 26' \,
57.6''$), followed by a telescope slew to the galaxy center (the dust
in NGC 7052 prevents a peak-up directly on the galaxy center
itself). Guided by simulations (van der Marel 1995), we used a
non-standard 5-stage peak-up sequence with a predicted accuracy (in a
noise-free situation) of $0.026''$ along each axis of the internal
$(X,Y)$ coordinate system of the FOS. The position of the galaxy
center with respect to the peak-up star was determined from the WFPC2
images before the spectroscopic observations: $\Delta {\rm RA} =
0.6130$s and $\Delta \delta = -7.796''$. Systematic errors in the
measurement of this offset and in the accuracy with which a slew of
this size can be performed are $\lta 0.01''$. As for the WFPC2
imaging, telescope tracking during the observations was done in `fine
lock', with a RMS telescope jitter of $\sim\! 3$ mas.

Galaxy spectra were obtained on the nucleus and along the major axis.
Target acquisition uncertainties and other possible systematic effects
may cause the aperture positions to differ from those commanded to the
telescope. We therefore determined the aperture positions from the
data themselves, using the target acquisition data, the ratios of the
continuum and emission-line flux as observed through different
apertures, and the FOS images obtained at the end of each visit. We
describe this analysis in Appendix~\ref{s:AppA}. Table~\ref{t:spectra}
lists the inferred aperture positions for all observations, as well as
the exposure times. We give the positions in an $(x,y)$ coordinate
system that is centered on the galaxy, and has its $x$-axis along the
galaxy major axis (sky position angle $63.5^{\circ}$). The aperture
positions are accurate to $\sim\!  0.02''$ in either
direction. Figure~\ref{f:aperpos} shows the aperture positions
overlaid on the WFPC2 {\II}-band image.

\placefigure{f:aperpos}
\placetable{t:spectra}

Most of the necessary data reduction steps are performed by the HST
calibration pipeline, including flat-fielding and absolute sensitivity
calibration. The wavelength calibration provided by the pipeline is
not accurate enough for our project, so we performed an improved
calibration using the arc lamp spectra obtained in each orbit,
following the procedure described in van der Marel (1997c). The
relative accuracy (between different observations) of the resulting
wavelength scale is $\sim 0.04${\AA} ($\sim\!2\kms$). Uncertainties in
the absolute wavelength scale are larger, $\sim\!0.4${\AA}
($\sim\!20\kms$), but influence only the systemic velocity of NGC
7052, not the inferred BH mass.

\subsection{Gas kinematics}
\label{ss:FOSgaskin}

The {\halnii} lines are the only emission lines with a
sufficiently high $S/N$ ratio for a kinematical analysis. To quantify
the gas kinematics we fit the spectra under the assumption that each
emission line is a Gaussian. The amplitude ratio H$\alpha$/[NII] is a
free parameter in each fit, but [NII]$6583/6548$ is always fixed to 3,
the ratio of the transition probabilities. We assume that all three
lines in a given spectrum have the same mean velocity $V$ and velocity
dispersion $\sigma$, and neglect the fact that the kinematics of the
H$\alpha$ and [NII] lines may be slightly different (as suggested by
Paper~I). Figure~\ref{f:emlines} shows the observed spectra and the
fits for all aperture positions.

\placefigure{f:emlines}

The observed emission lines are not perfectly fit by Gaussians: they
have a narrower core and broader wings. We discuss the line shapes in
Section~\ref{ss:results}. For now we restrict the discussion to the
Gaussian fit parameters $V$ and $\sigma$, which are listed with their
formal errors in Table~\ref{t:spectra} (note that the mean and
dispersion of the best-fitting Gaussian are well-defined and
meaningful quantities, even if the lines themselves are not
Gaussians).  The systemic velocity of NGC 7052 (which was subtracted
from the velocities in Table~\ref{t:spectra}) was determined from the
observed mean velocities by including it as a free parameter in the
models discussed below, which yields $v_{\rm sys} = 4705 \pm 10 \kms$.
This is consistent with the velocity $v_{\rm sys} = 4675 \pm 45 \kms$
inferred from the absorption lines in the HST spectra, and also with
the literature value of $v_{\rm sys} = 4710 \pm 36 \kms$ (Di Nella
\etal 1995).

Figure~\ref{f:gaskin} shows the kinematical quantities inferred from
the FOS data as function of major axis distance. The figure also shows
the kinematics inferred (in similar fashion using single-Gaussian
fits) from the major axis ground-based spectra presented in
Paper~I. The line widths increase strongly towards the nucleus,
similar to what has been found for other galaxies with nuclear gas
disks (e.g., Ferrarese, Ford \& Jaffe, 1996; Macchetto \etal 1997;
Bower \etal 1998). As compared to the ground-based NGC 7052 data, the
rotation curve inferred from the HST data is twice as steep, and the
emission lines observed on the nucleus are twice as broad. To
interpret these results we construct dynamical models for the gas
motions.

\placefigure{f:gaskin}

\section{The nuclear mass distribution}
\label{s:dyn}

\subsection{Dynamical models}
\label{ss:models}

Our models for the gas kinematics are similar to those employed in
Paper~I. The galaxy model is axisymmetric, with the stellar luminosity
density $j(R,z)$ chosen as in Section~\ref{ss:WFduststars} to fit the
available surface photometry. We use our standard model for $j(R,z)$,
unless otherwise specified. The stellar mass density $\rho(R,z)$
follows upon specification of a mass-to-light ratio $\Upsilon$. The
gas resides in an infinitesimally thin disk in the equatorial plane of
the galaxy, and has the circularly symmetric flux distribution $F(R)$
given in Section~\ref{ss:WFgas}. The galaxy and the gas disk are
viewed at an inclination $i$.  The mean motion of the gas is
characterized by circular orbits. The circular velocity $\Vc(R)$ is
calculated from the combined gravitational potential of the stars and
a nuclear BH of mass $\Mbh$.  The line-of-sight velocity profile (VP)
of the gas at position $(x,y)$ on the sky is a Gaussian with
dispersion $\sigma_{\rm gas}(R)$ and mean $\Vc(R) \sin i$, where $R^2
= x^2 + (y/\cos i)^2$ is the radius in the disk. The velocity
dispersion of the gas is assumed to be isotropic, with contributions
from thermal and non-thermal motions: $\sigma_{\rm gas}^2 =
\sigma_{\rm th}^2 + \sigma_{\rm turb}^2$. We refer to the non-thermal
contribution as `turbulent', although we make no attempt to describe
the underlying physical processes. It was concluded in Paper~I that
the intrinsic dispersion of the gas in NGC 7052 increases towards the
nucleus, and it is sufficient here to parameterize $\sigma_{\rm turb}$
through:
\begin{equation}
  \sigma_{\rm turb}(R) = \sigma_0 + [\sigma_1 \exp(-R/\Rt)] ,
\label{eq:turbdef}
\end{equation}
where $R$ is the radius in the disk. The predicted VP for any given
observation is obtained through flux weighted convolution of the
intrinsic VPs with the PSF of the observation and the size of the
aperture. The convolutions are described by the semi-analytical
kernels given in Appendix~A of van der Marel \etal (1997b), and were
performed numerically using Gauss-Legendre integration. A Gaussian is
fit to each predicted VP for comparison to the observed $V$ and
$\sigma$.

In a spectrograph, light at different positions in the aperture is
detected at slightly different wavelengths. This induces line
broadening, as well as small velocity shifts if the light is not
distributed symmetrically within the aperture. Instrumental broadening
also results from the finite size of a detector resolution element,
and from the broadening due to the grating itself. For each
observation we calculated the instrumental line spread function due to
these effects as in Appendix~B of van der Marel \etal (1997b). The
resulting corrections on the predicted VPs and kinematical quantities
were included in the models, but are not large ($\lesssim 6 \kms$ for the
inferred mean velocities).

Five free parameters are available to optimize the fit to the observed
gas kinematics: $\Mbh$, $\Upsilon$, and the parameters $\sigma_0$,
$\sigma_1$ and $\Rt$ that describe the radial dependence of the
turbulent dispersion. The temperature of the gas is not an important
parameter: the thermal dispersion for $T \approx 10^4 {\rm K}$ is
$\sigma_{\rm th} \approx 10\kms$, and is negligible with respect to
$\sigma_{\rm turb}$ for all plausible models. We define a $\chi^2$
quantity that measures the quality of the fit to the kinematical
data. The best-fitting model is found by minimizing $\chi^2$ using a
`downhill simplex' minimization routine (Press \etal 1992).

Figure~\ref{f:modelfits} shows the data that we have used in the
definition of $\chi^2$, namely the new FOS data and the ground-based
major and minor axis WHT data presented in Paper~I. The latter were
obtained in FWHM seeing of $0.57''$ and $0.84''$, respectively.
Rotation velocity and velocity dispersion measurements were both
included in the fit, yielding a total of 44 data points. The WHT data
show some features that are in contradiction with the assumption of
perfect axisymmetry: the velocity dispersion profiles along both the
major and minor axes are not quite symmetric, and there is a small
amount of rotation along the minor axis. This implies that the models
can never fit the WHT measurements to within their formal errors,
which are significantly smaller than those for the FOS data (a result
of higher $S/N$). However, the models will automatically search for
the best fit to the average of the measurements at positive and
negative radii. We do not believe that this compromises our ability to
infer the nuclear mass distribution of NGC 7052 with axisymmetric
models. However, one complication is the weighting of data with its
formal errors in the usual definition of $\chi^2$. This causes the
minimization routine to assign disproportionate priority to features
in the WHT data that it can never fit, while neglecting the fit to the
FOS data. This is not desirable, because the FOS data have higher
spatial resolution, and therefore contain more information on the mass
distribution at small radii. To ensure that the FOS and WHT data
receive roughly equal weight in the fitting process, we artificially
assigned all WHT data points an error of $22.5 \kms$ in the definition
of $\chi^2$, equal to the median error for the HST data points.
Although somewhat artificial, we have found that this works well in
practice. It does mean that the resulting $\chi^2$ cannot be used for
the calculation of confidence intervals on the best-fitting model
parameters. We realize this, and therefore calculate in
Section~\ref{ss:Mbhrange} confidence intervals on $\Mbh$ on the basis
of a statistic that only uses the FOS observations (for which the
errors were taken as observed).

\placefigure{f:modelfits}

\subsection{The best-fit model}
\label{ss:results}

The curves in Figure~\ref{f:modelfits} show the predictions of the
model that provides the overall best fit to the data. Its parameters
are: $\Mbh = 3.3 \times 10^8 \Msun$, $\Upsilon = 6.3$ (in {\II}-band
solar units), $\sigma_0 = 60 \kms$, $\sigma_1 = 523 \kms$ and $\Rt =
0.11''$. This model adequately reproduces all the important features
of the kinematical data, including the rotation curve slope and the
nuclear velocity dispersion inferred from the HST data. The fit of our
flux distribution model (equation~[\ref{fluxparam}]) to the observed
fluxes (top panel of Figure~\ref{f:modelfits}) was discussed already
in Section~\ref{ss:WFgas}.

The circular velocity curve $\Vc(R)$ of a model with a BH has a
minimum that is non-zero (for our best-fit model the minimum of
$\Vc(R) \sin i$ is $180 \kms$). Intrinsically, no gas moves slower
than this, although gas for which only a fraction of the orbital
velocity is observed along the line of sight may be observed at lower
velocities. Hence, there is a tendency for models with a BH to predict
double-peaked line profiles for observations on the nucleus, with the
peaks at $|v| \approx [\Vc(R)]_{\rm min} \sin i$, corresponding to gas
seen moving towards and away from the observer on either side of the
nucleus. However, a significant intrinsic dispersion for the gas will
tend to `wash away' the peaks. Our best-fit model has both a BH and a
significant intrinsic velocity dispersion. It is therefore interesting
to study whether its predicted line shapes are comparable to the
observed ones. Figure~\ref{f:lineshapes}b compares the predicted
spectrum to the data for FOS observation \#4 (see
Table~\ref{t:spectra}), which was obtained on the nucleus. The model
does not predict double-peaked or flat-topped line profiles, but
instead predicts profiles with a narrow core and broad wings, as seen
in the observations. The predictions are acceptable, and fit better
than Gaussian profiles (shown in Figure~\ref{f:lineshapes}a for
comparison). Similar results were obtained for the other FOS
observations. The fits to the line shapes could conceivably be
improved further by allowing the intrinsic velocity distribution in
the disk at any given position to be non-Gaussian, but this is not
something that we have explored. It is already clear that our best-fit
model is consistent with the observed line shapes.

\placefigure{f:lineshapes}

The observed velocity dispersion of the gas increases from
$\sim\!70\kms$ at $1''$ from the nucleus, to $\sim\!400\kms$ as seen
through the $0.26''$ diameter FOS aperture on the nucleus. In
principle, one can get an increase in the observed velocity dispersion
towards the nucleus even if there is no intrinsic gradient, through
the effect of rotational broadening around a central BH. However, it
was shown already in Paper~I that a model with a BH and no intrinsic
velocity dispersion gradient is unacceptable: such a model predicts
double-peaked emission line shapes that are in contradiction with
observations. We similarly find that such models fail to fit the HST
data. Hence, the intrinsic velocity dispersion of the gas in NGC 7052
must increase towards the nucleus. Our best-fit model has a turbulent
velocity dispersion that increases to values in excess of $500\kms$ in
the central $0.1''$. We argue in Section~\ref{ss:starkin} that this is
not in contradiction with the assumption of bulk circular motion for
the gas. Thus, the ionized gas in NGC 7052 has both a sharply
increasing flux and a sharply increasing velocity dispersion towards
the nucleus. It may be that this signifies the presence of a (partly
unresolved) narrow-line region. There is no evidence for a broad-line
region, because the increasing line width is not restricted to
H$\alpha$, but is seen in the forbidden [NII] lines as~well.

\subsection{The allowed range of black hole masses}
\label{ss:Mbhrange}

To determine the range of BH masses that provides a statistically
acceptable fit to the data, we compare the predictions of models with
different, fixed values of $\Mbh$, but in which the remaining
parameters are varied to optimize the fit. The radial dependence of
the intrinsic velocity dispersion of the gas is essentially a free
function in our models, so the observed velocity dispersion
measurements can be fit equally well for all plausible values of
$\Mbh$. The observed ground-based rotation measurements can also be
fit equally well for all plausible $\Mbh$ (i.e., $\lesssim 10^9
\Msun$; cf.~Paper~I), because for all such values the radius of the BH
sphere of influence is smaller than the ground-based spatial
resolution.  Thus only the predictions for the HST rotation velocity
measurements depend substantially on the adopted $\Mbh$.

Figure~\ref{f:rotfits} compares the predictions for the HST rotation
measurements for three different models. The solid curve is the
best-fit model. The dashed curves are models in which $\Mbh$ was fixed
a priori to $0$ and $1.0 \times 10^9 \Msun$, respectively. The error
weighted mean of the three velocity observations at $0.2''$ from the
center is $V = 155 \pm 17 \kms$. The model without a BH predicts $V =
92 \kms$, and its rotation curve slope is thus too shallow.  The model
with $\Mbh = 1.0 \times 10^9 \Msun$ predicts $V = 214\kms$, and its
rotation curve slope is thus too steep.

\placefigure{f:rotfits}

To assess the quality of the fit to the HST rotation velocity
measurements we define a new $\chi^2$ quantity, $\chi^2_V$, that
measures the fit to these data only. Figure~\ref{f:chisqv} shows
$\chi^2_V$ as function of $\Mbh$. At each $\Mbh$, the parameters
$\Upsilon$, $\sigma_0$, $\sigma_1$ and $\Rt$ are fixed almost entirely
by the ground-based data and the HST velocity dispersion measurements.
These parameters can therefore not be varied independently to improve
the fit to the HST rotation velocity measurements. As a result,
$\chi^2_V$ follows approximately a $\chi^2$ probability distribution
with $N_{\rm df} = 6-1 = 5$ degrees of freedom (there are six HST
measurements, and there is one free parameter: $\Mbh$). The
expectation value for this distribution is $\langle \chi^2_V
\rangle = 5$. The best-fit model has $[\chi^2_V]_{\rm min} =
3.8$, and is thus entirely consistent with the data. Horizontal lines
in Figure~\ref{f:chisqv} show the probability that a value equal to or
smaller than indicated would be observed for a correct model. These
lines show that the $68.3$\% (i.e., 1-$\sigma$) confidence interval
for $\Mbh$ is $[2.5\,;\,5.0] \times 10^8 \Msun$, the $90$\% confidence
interval is $[1.9\,;\,6.5] \times 10^8 \Msun$, and the $99$\%
confidence interval is $[1.2\,;\,8.2] \times 10^8\Msun$. Models
without a BH are ruled out at $>99$\% confidence.

\placefigure{f:chisqv}

\subsection{Dependence on cusp slope}
\label{ss:nonstandard}

In Sections~\ref{ss:results} and~\ref{ss:Mbhrange} we analyzed the gas
kinematics using our standard model for the stellar luminosity
density, which has a central cusp slope $\alpha=0.5$
(cf.~Section~\ref{ss:WFduststars}). To assess the dependence of the
inferred BH mass on $\alpha$ we repeated the analysis for a range of
other $\alpha$ values, namely $\alpha=0.0,0.1,0.2,0.3,\ldots,1.3$. 

Changing $\alpha$ from its standard value leaves the quality of the
fit to the gas kinematics virtually unchanged; when $\alpha$ is
changed, $\Mbh$ and $\Upsilon$ can be changed simultaneously in such a
way as to maintain a similar circular velocity curve. The best-fit
dynamical models for different values of $\alpha$ therefore all have a
similar turbulent velocity dispersion profile, but different values of
$\Upsilon$ and $\Mbh$. The dependence of the mass-to-light ratio on
$\alpha$ is not very strong; $\Upsilon$ decreases monotonically with
$\alpha$, from $\Upsilon = 6.8$ for $\alpha=0.0$ to $\Upsilon = 5.2$
for $\alpha=1.3$. The dependence of $\Mbh$ on $\alpha$ is more
important in the present context, and is shown in
Figure~\ref{f:alphadep}. This figure not only shows the best-fit
$\Mbh$ for each $\alpha$, but also the $68.3$\%, $90$\% and $99$\%
confidence intervals, determined using the approach of
Section~\ref{ss:Mbhrange}. The best-fit value of $\Mbh$ decreases
monotonically with $\alpha$, from $\Mbh = 3.9 \times 10^8 \Msun$ for
$\alpha=0.0$ to $\Mbh = 1.8 \times 10^8 \Msun$ for
$\alpha=1.3$. Models with larger values of $\alpha$ have more stellar
mass near the center, and therefore do not need as large a BH mass to
explain the observed rotation curve.  However, the best-fit model does
invoke a BH for all values of $\alpha$. Models without a BH always
predict a rotation curve that is shallower than observed, and are
ruled out with $>99$\% confidence over the entire range $0.0 \leq
\alpha \leq 1.3$.

\placefigure{f:alphadep}

Independent constraints on $\alpha$ were presented in
Section~\ref{ss:WFduststars}. If NGC 7052 follows the trends displayed
by other elliptical galaxies, then its luminosity determines that it
must be a core galaxy. This yields $\alpha \lta 1.0$. Furthermore, the
observed {\BB}$-${\II} color of the dust obscured region implies that
$\alpha \lta 0.8$.\footnote{These arguments rule out the value $\alpha
= 1.3$ that we adopted in Paper~I on the basis of more limited models
for ground-based data.} If we assume that $\alpha = 0.5$ is the most
likely value for $\alpha$ (based on Figure~\ref{f:reddening}), but
that none of the values of $\alpha$ in the range $0.0 \leq \alpha \lta
0.8$ can be ruled out, we obtain $\Mbh = 3.3^{+2.3}_{-1.3} \times 10^8
\Msun$ at $68.3$\% confidence.

\subsection{Stellar kinematics} 
\label{ss:starkin}

In Paper~I we presented a single measurement $\sigma_{\ast} = 266 \pm
26 \kms$ of the stellar velocity dispersion in NGC 7052, obtained from
the summed spectrum over a $10''$ wide and $0.8''$ high region along
the major axis. To interpret this measurement we constructed stellar
dynamical models with a phase-space distribution function of the form
$f(E,L_z)$ (as in Paper~I), for $\alpha=0.0,0.1,0.2,0.3,\ldots,1.3$,
respectively. For each $\alpha$, the values of $\Upsilon$ and $\Mbh$
were taken to be as determined by the best fit to the gas kinematics
(as in Sections~\ref{ss:results} to~\ref{ss:nonstandard}). The stellar
dynamics of each model were projected onto the sky and binned to
obtain a prediction $\sigma_{\rm pred}$ for the
observed~$\sigma_{\ast}$.

The predicted dispersion decreases monotonically from $\sigma_{\rm
pred} = 272 \kms$ for $\alpha=0.0$ to $\sigma_{\rm pred} = 245 \kms$
for $\alpha=1.3$. This implies that the predictions are consistent
with the data for all values of $\alpha$. This is an important
consistency check on our models for the gas kinematics. In particular,
it confirms that the assumption of circular orbits for the gas is
reasonable. If the gas in NGC 7052 were to move at significantly
sub-circular velocities, then our models for the gas kinematics would
yield an underestimate of the true mass-to-light $\Upsilon$. When used
in our stellar dynamical models, this anomalously low $\Upsilon$ would
lead to an underprediction of the observed stellar velocity
dispersion, which we do not find to be the case.

This result lends support to our treatment of the intrinsic velocity
dispersion of the gas. In our models we assume that this dispersion is
due to local turbulence in gas that has bulk motion along circular
orbits. An alternative would be to assume that the gas resides in
individual clouds, and that the observed dispersion of the gas is due
to a spread in the velocities of individual clouds seen along the
line-of-sight. However, in this scenario the gas would behave
similarly as other point particles (e.g., stars), and would therefore
rotate significantly slower than the circular velocity (due to
asymmetric drift). Our models would then have produced an
underestimate of $\Upsilon$, and with this $\Upsilon$ our stellar
dynamical models would have underpredicted the observed stellar
velocity dispersion. As mentioned, we do not find this to be the case.
This adds to the fact that the inferred $V/\sigma$ for the gas is less
than unity. If the observed gas dispersion were due to gravitational
motion of individual clouds, this $V/\sigma$ would be in contradiction
with the fact that both the gas and the dust appear to reside in flat
disks.

\subsection{Adiabatic black hole growth}
\label{ss:adiab}

HST observations of early-type galaxies have shown that central
power-law surface brightness cusps are ubiquitous. One possible
scenario for the formation of such cusps is through adiabatic BH
growth into a pre-existing homogeneous core (Young 1980). Although
this is not a unique explanation, it may well be the correct one. For
the case of M87 it has been shown to be in perfect agreement with the
dynamically determined BH mass (Young \etal 1978; Lauer \etal
1992). Furthermore, the assumption that the cusps in all core galaxies
are due to adiabatic BH growth implies a BH mass distribution that is
in excellent agreement both with results from well-studied individual
galaxies and with predictions from quasar statistics (van der Marel
1998).

If we assume that the presence of a surface brightness cusp of slope
$\alpha$ in NGC 7052 is due to adiabatic BH growth as envisaged by
Young, then each value of $\alpha$ corresponds to a unique BH mass
$\Mbh$. We determined the dependence of $\Mbh$ on $\alpha$ by fitting
adiabatic BH growth models calculated with the software of Quinlan
\etal (1995) to the surface brightness profiles that correspond to
the luminosity density parameterization of equation~(\ref{lumdendef}).
The result is shown in Figure~\ref{f:alphadep}. The BH mass thus
inferred agrees at the $90$\% confidence level with the BH mass
inferred from the gas kinematics if $0.16 \leq \alpha \leq 0.53$.
This includes the value $\alpha=0.5$ of our standard model.

These results provide new evidence for the applicability of Young's
models to core galaxies. It should be noted though that this does not
necessarily imply that the BHs in core galaxies grew adiabatically. If
the BH was present even before the galaxy formed, the end product
would be very similar (Stiavelli 1998).

\section{Discussion and conclusions}
\label{s:disc}

We have presented HST observations of the nuclear gas and dust disk in
the E4 radio galaxy NGC 7052. WFPC2 broad- and narrow-band images were
used to constrain the stellar surface brightness profile, the optical
depth of the dust, and the flux distribution of the ionized gas. We
have built axisymmetric models in which the gas and dust reside in the
equatorial plane, and in which the gas moves on circular orbits with
an additional velocity dispersion due to turbulence (or otherwise
non-gravitational motion). These models were used to interpret the
ionized gas kinematics inferred from our new FOS spectra and from
existing ground-based spectra. The models fit the observed central
rotation gradient only if there is a central BH with mass $\Mbh =
3.3^{+2.3}_{-1.3} \times 10^8 \Msun$. Models without a black hole are
ruled out at $>99$\% confidence.

The models provide an adequate fit to the available observations with
a minimum number of free parameters. The assumptions that we make are
similar to those that have been made in HST studies of other galaxies
with nuclear gas disks. In several areas our models are in fact more
sophisticated than some of the previous work. In particular: we use
our multi-colour photometry in order to constrain the central cusp
steepness of the stellar mass distribution; we explicitly take into
account the contribution of the axisymmetric stellar mass distribution
to the circular velocity of the gas, and we do not assume the rotation
field to be purely Keplerian; we explicitly model the convolution with
the HST/FOS PSF and the binning over the size of the aperture; we
model the full line profile shapes, and fit the widths of the emission
lines as well as their mean; and we fit Gaussians to the models as we
do the data, to properly take into account the fact that Gaussian fits
to lines that may be skewed or have broad wings yield biased estimate
of the true moments.

Still, our models remain only an approximation to the true structure
of NGC 7052. In particular: the thickness of the gas disk may not be
negligible; the mean motion of the gas may not be circular; and the
observed rotation curve may not perfectly reflect the intrinsic
rotation curve, because of partial absorption of the emission line
flux by dust. The limited sky coverage of the FOS spectra prevents a
direct check on whether the gas motions in NGC 7052 are indeed
circular. However, several consistency checks are available that may
have signaled errors in our assumptions; none did. The stellar
mass-to-light ratio and systemic velocity inferred with our models
from the nuclear gas kinematics agree with those inferred from stellar
kinematical measurements outside the region influenced by dust
absorption. The best-fitting model for the gas kinematics reproduces
the shapes of the emission lines on the nucleus, despite the fact that
these shapes were not included as constraints in the fit. These
agreements do not rule out a conspiracy of some sort, but they do make
it less likely that the observed gas kinematics are the result of
vastly non-circular motion, or have been strongly modified by dust
absorption. Models of adiabatic BH growth for the stellar surface
brightness cusp provide another successful check: the BH mass implied
by these models is fully consistent with that inferred from the gas
kinematics.

Figure~\ref{f:allBHs} shows a scatter plot of $\Mbh$ versus {\BB}-band
spheroid luminosity $L_{B,{\rm sph}}$ for all galaxies with reasonably
secure BH mass determinations (adapted from van der Marel 1998, with
the addition of NGC 7052; all for $H_0 = 80 \kms \Mpc^{-1}$). There is
a trend of increasing $\Mbh$ with increasing $L_{B,{\rm sph}}$,
although it remains difficult to rule out that systematic biases play
some role in this relation (van der Marel 1998). Besides NGC 7052, the
other galaxies for which the BH detections are based on kinematical
studies of nuclear gas disks with the HST are M87 (Harms \etal 1994;
Macchetto \etal 1997), M84 (Bower \etal 1998), NGC 6251 and NGC 4261
(Ferrarese, Ford \& Jaffe, 1998, 1996). The $\Mbh$ in these galaxies
are $3.2 \times 10^9$, $1.4 \times 10^9$, $6.6 \times 10^8$ and $\Mbh
= 4.9 \times 10^8 \Msun$, respectively. NGC 7052 falls at the low end
of this range. The five galaxies with BH evidence from nuclear gas
disks form a very homogeneous set. Each of these galaxies is a radio
source and is morphologically classified as an elliptical. The
luminosities are identical to within $\sim\!25$\% ($\log L_B$ in the
range $10.6$---$10.8$ for all five galaxies). By contrast, the black
hole masses span a range of a factor $10$. The results for these
galaxies therefore show that any relation between $\Mbh$ and
$L_{B,{\rm sph}}$ must have a scatter of at least a factor $10$, even
if the comparison is restricted to galaxies of similar type.

\placefigure{f:allBHs}


\acknowledgments

The authors are grateful to Tim de Zeeuw and Tony Keyes for helpful
discussions, and to Bill Workman and Jean Surdej for successful
scheduling and implementation of the observations. Gerry Quinlan
kindly allowed us to use his adiabatic BH growth software. FvdB thanks
the Institute for Advanced Study in Princeton for its hospitality
during a visit in 1996. Support for this work was provided by NASA
through grant number \#GO-05848.01-94A, through Hubble Fellowships
\#HF-01065.01-94A and \#HF-01102.11-97A, and through an STScI Fellowship, 
all awarded by the Space Telescope Science Institute which is operated
by the Association of Universities for Research in Astronomy,
Incorporated, under NASA contract NAS5-26555.


\clearpage

\appendix

\section{Aperture positions for the FOS observations}
\label{s:AppA}

In the first FOS visit (September 1995) the instrument was commanded
to obtain spectra on the nucleus and at $\pm 0.2''$ along the major
axis, and in the second visit (August 1996) on the nucleus and at $\pm
0.16''$ along the major axis.  To determine the actual aperture
positions during the observations we make the simplifying assumption
that the relative positions of the apertures during each visit were as
intended, but that for each visit there is one absolute offset
(corresponding to the difference between the actual position of the
galaxy center and the telescope's estimate of the galaxy center). This
assumption is justified if there is little telescope drift during the
observations (telescope drifts during a visit are usually small,
$\lesssim 0.03''$, although larger drifts can sometimes occur; e.g.,
van der Marel \etal 1997b). We have determined the absolute offset for
each visit using three different methods, based on: (i) the target
acquisition data; (ii) the ratios of either the continuum or the
emission-line intensity between observations at different positions;
and (iii) the FOS image obtained at the end of each visit.

The final target acquisition stage on the peak-up star placed the {\tt
0.3} aperture at a $5 \times 5$ grid on the sky with inter-point
spacings of $0.052''$. The intensity at each position was measured,
and the grid point with the highest intensity was adopted by the
telescope as its estimate of the position of the star. The intensities
for all 25 positions were returned to the ground, and could be
interpolated {\it post hoc} to determine the actual position of the
star, which is not generally exactly at a grid point. This yields a
direct estimate of the absolute positional offset for each visit,
under the assumption that target acquisition inaccuracies provide the
dominant source of positional error.

Each galaxy spectrum yields the total intensity in the {\halnii}
lines and in the surrounding continuum. We binned the intensities
observed in the WFPC2 on-band and off-band images in $0.26''$ circular
apertures, and determined the absolute positional offset that best
reproduces the intensities observed in the spectra. This yields a
second estimate of the absolute positional offset for each visit, but
only for the component along the major axis (both the intensity
distribution and the positioning of the apertures are aligned along
the major axis, so any positional offset in the direction parallel to
the minor axis yields little change in the predicted intensity
ratios).

The FOS image obtained at the end of each visit provides a third
estimate of the absolute positional offset. The telescope was
commanded to position the {\tt 4.3} aperture (a square aperture of
$3.66'' \times 3.71''$) at its estimate of the galaxy center, and to
record an image by stepping the diode array of the instrument along
the aperture. This yields an image with $0.075'' \times 0.081''$
pixels, with a point-spread-function equal to the size of one diode
($0.301'' \times 1.291''$). We modeled the image for each visit by
masking the WFPC2 {\II}-band image with the aperture size, convolving it
with the boxcar PSF of one diode, and determining the absolute
positional offset that provides the best fit in a $\chi^2$ sense to
the observed FOS image.

For each visit, the three different methods for determining the
absolute positional offset yield results that are mutually consistent
to within $0.02''$ in each coordinate. For the first spectroscopic
visit we find $(\Delta x,\Delta y) = (0.00'',0.01'')$, and for the
second visit $(\Delta x,\Delta y) = (-0.04'',-0.02'')$, where $(x,y)$
is the coordinate system used in Section~\ref{ss:FOSobsetup}.

\clearpage


\ifsubmode\else
\baselineskip=10pt
\fi


\clearpage


\ifsubmode\else
\baselineskip=14pt
\fi


\newcommand{\figcapimages}{WFPC2 images of NGC 7052. From top-left to bottom
  right: {\II}-band; {\BB}-band; $B-I$; and {\halnii}. Each panel
  covers a square region of $7.75''$. The {\II}, {\BB}, and $B-I$
  image were taken with the PC chip, yielding $0.046''$ pixels, and
  are shown with a linear strech.  The darkest regions in the $B-I$
  image have $B-I \approx 3.0$.  The stellar distribution outside the
  region influenced by the disk has $B-I \approx 2.5$. The emission
  line image was constructed from narrow-band images taken with the WF
  chips, yielding $0.010''$ pixels. It has a sharply peaked flux
  distribution, but is shown here with a logarithmic stretch to show
  the low surface brightness features. The orientation on the sky and
  the position angle of the radio jet are indicated in the $B-I$
  image.  The regions labeled~A and~B are discussed in the text.
  \label{f:images}}

\newcommand{\figcapellipsefit}{Isophotal parameters as function of major
  axis radius for the $I$-band image of NGC~7052, for the region
  outside the dust disk ($r > 2.0''$). From left to right and top to
  bottom: the $I$-band surface brightness in mag arcsec$^{-2}$; the
  $B-V$ and $V-I$ color profiles; the ellipticity $\epsilon$; the
  major axis position angle $\theta$; and the third and fourth order
  Fourier sine ($A_3$ and $A_4$) and cosine ($B_3$ and $B_4$) terms
  that measure isophotal deviations from ellipses.  The isophotal
  structure of NGC~7052 is not atypical for an elliptical
  galaxy.\label{f:ellipsefit}}

\newcommand{\figcapminorprof}{The heavy solid line in the top panel is the
  observed {\II}-band intensity profile along the minor axis, averaged
  over a $0.14''$ wide strip centered on the galaxy. The abscissa is
  the distance to the galaxy center, measured along the minor axis in
  the direction least affected by dust obscuration (`upward' in
  Figure~\ref{f:images}). The dashed vertical line indicates the
  approximate extent of the dust disk.  Dotted curves are predictions
  of models with the stellar luminosity density parameterization of
  equation~(\ref{lumdendef}), for $\alpha=0.1$, $0.2$, $0.3, \ldots,
  1.3$. Our standard model has $\alpha=0.5$, and is shown with a solid
  curve. At each fixed $\alpha$, the remaining model parameters were
  chosen to optimize the fit to the data outside the dust region (to
  the right of the dashed line). The models take PSF convolution and
  pixel binning into account. The bottom panel shows the residuals of
  each of the fits.\label{f:minorprof}}

\newcommand{\figcapminorcolor}{The $B-I$ color as function of minor axis
  distance, averaged over a $0.14''$ wide strip centered on the
  galaxy. Positive radii correspond to the upward direction in
  Figure~\ref{f:images}. Dotted vertical lines indicate the
  approximate extent of the dust disk. The color variation outside the
  central arcsec is due to a radial stellar population gradient. The
  solid line provides a linear fit to this gradient, and is
  extrapolated into the disk region as a dashed line.  The disk has
  its largest $B-I$ at negative radii, at the front side of the disk.
  The nucleus is slightly bluer than its surroundings, possibly due to
  a contribution from non-thermal continuum.\label{f:minorcolor}}

\newcommand{\figcapreddening}{The left panel shows $\chi^2_{\alpha}$
  as function of the intrinsic central cusp slope $\alpha$. The
  quantity $\chi^2_{\alpha}$ measures the quality of the fit to the
  observed $B$-band flux for the predictions obtained as described in
  the text.  In essence, $\alpha$ fixes the optical depth of the dust
  in the {\II}-band, and $\chi^2_{\alpha}$ measures how well this
  optical depth fits the observed reddening of the galaxy light. The
  right panel shows the fraction $f$ of pixels that has a smaller
  {\BB}-band flux than would be expected for infinite optical
  depth. Small values of $f$ can be attributed to Poisson noise, but
  large values indicate that the underlying model is incorrect. We
  conclude from the combined results in both panels that $\alpha \lta
  0.8$ with a most likely value of $\alpha = 0.5$.
  \label{f:reddening}}

\newcommand{\figcapcentralslope}{The NUKER law fit parameter $\gamma$ as
  function of the parameter $\alpha$ of our models, when fit over the
  region between $0.1''$ and $10''$. The former measures the central
  logarithmic slope of the projected intensity, the latter the central
  logarithmic slope of the luminosity density. At asymptotically small
  radii the two are related according to $\gamma = \max(0,\alpha-1)$,
  as indicated by the dashed line.\label{f:centralslope}}

\newcommand{\figcapgasfit}{Solid dots show the {\halnii} flux (in arbitrary 
  units) as function of the distance $R$ from the center, measured
  from the WFPC2 emission-line image along the major axis, the
  intermediate axis, and the minor axis, respectively. The dotted
  curve in the left panel shows the WFPC2 PSF. The observed flux
  distribution is peaked towards the center, but is broader than the
  PSF, and thus resolved. The solid curves in each panel show the
  predictions of the best-fitting model with the flux distribution of
  equation~(\ref{fluxparam}), after convolution with the PSF and pixel
  size. The parameters of this model were chosen to simultaneously
  optimize the fit to the fluxes obtained from the WFPC2 data and
  those obtained from the FOS and ground-based spectra (see the top
  row of Figure~\ref{f:modelfits}). The flux distribution outside the
  central region is somewhat asymmetric, in particular along the major
  axis. This is not reproduced by the model, but the overall fit is
  satisfactory.\label{f:gasfit}}

\newcommand{\figcapaperpos}{Aperture positions for the HST/FOS spectra, 
  overlaid on the HST/WFPC2 {\II}-band image of NGC 7052. All
  observations were obtained with the circular FOS {\tt 0.3} aperture,
  which has a diameter of $0.26''$. Table~\ref{t:spectra} lists
  various pieces of information about the spectra. The orientation of
  the image is the same as in Figure~\ref{f:images}.\label{f:aperpos}}

\newcommand{\figcapemlines}{Dotted lines show the continuum-subtracted
  {\halnii} emission lines in the HST/FOS spectra of NGC 7052. The
  instrumental resolution is $4.15${\AA} FWHM, corresponding to a
  dispersion (FWHM/$2.355$) of $79 \kms$. The spectra were smoothed
  with a Gaussian with a dispersion of $1${\AA} to reduce the noise.
  The abscissa is the observed vacuum wavelength in {\AA}, and the
  ordinate is the flux observed through the $0.26''$ diameter aperture
  in $10^{-16}$ erg cm$^{-2}$ s$^{-1}$ {\AA}$^{-1}$. A label in each
  panel indicates both the number of the observation, as defined in
  Table~\ref{t:spectra}, and the major axis distance $x$ of the
  aperture center. Heavy solid curves are the fits to the spectra
  obtained under the assumption that all emission lines in a given
  spectrum are Gaussians with the same mean $V$ and dispersion
  $\sigma$, as described in the text. The individual Gaussian fits to
  [NII]$6548$, H$\alpha$ and [NII]$6583$ are displayed with a negative
  offset. The kinematical quantities $V$ and $\sigma$ are shown in
  Figure~\ref{f:gaskin}.\label{f:emlines}}

\newcommand{\figcapgaskin}{Mean velocity $V$ and velocity dispersion $\sigma$
  in $\kms$ as function of major axis distance $x$ in arcsec, for
  Gaussian fits to the {\halnii} emission lines in NGC 7052. The solid
  points are the results obtained with the HST/FOS from observations
  with a $0.26''$ diameter aperture, as listed in
  Table~\ref{t:spectra}. The actual fits to the spectra are shown in
  Figure~\ref{f:emlines}. The connected symbols are the results
  obtained from the major axis ground-based WHT spectra presented in
  Paper~I, obtained in $0.57''$ FWHM seeing. The kinematical gradients
  inferred from the higher spatial resolution HST data are
  significantly steeper.\label{f:gaskin}}

\newcommand{\figcapmodelfits}{Data-model comparison for the flux and kinematics
  of the ionized gas. Gaussian fits to the emission lines were used to
  obtain measurements of the total {\halnii} flux $I$ (in $10^{-14}$
  erg cm$^{-2}$ s$^{-1}$ per aperture), the mean velocity $V$, and the
  velocity dispersion $\sigma$ (in $\kms$). From left to right, the
  observations are from the HST/FOS, with $0.26''$ diameter apertures
  aligned along the major axis, and from long-slit ground-based WHT
  observations along the major and minor axes, obtained with seeing
  FWHM of $0.57''$ and $0.84''$, respectively, and with $0.33'' \times
  0.8''$ apertures. The horizontal scale in the panels for the HST
  data is five times smaller than in the other panels. Errors for the
  WHT data are often smaller than the plot symbols. The curves in the
  top row are the predictions generated by our model for the intrinsic
  flux distribution of the gas (equation~[\ref{fluxparam}]). The
  curves in the bottom two rows are the predictions of our best-fit
  model for the gas kinematics, using our standard model for the
  stellar luminosity density ($\alpha = 0.5$). The dynamical model has
  a BH of $3.3 \times 10^8 \Msun$, and provides an adequate
  fit.\label{f:modelfits}}

\newcommand{\figcaplineshapes}{The dotted line in each panel is the
  continuum-subtracted HST/FOS spectrum for observation \#4 (see
  Table~\ref{t:spectra}), which was obtained on the nucleus. The top
  panel shows the best-fit when each emission line is assumed to have
  a Gaussian shape (as in Figure~\ref{f:emlines}). The individual fits
  to [NII]$6548$, H$\alpha$ and [NII]$6583$ are displayed with a
  negative offset. The bottom panel shows the fit to the spectrum when
  the shape and width of each line are chosen as predicted by the
  best-fit model (Figure~\ref{f:modelfits}), with the flux in each of
  the lines and the mean velocity chosen to optimize the fit to the
  data. The predicted line shapes have a narrower core and broader
  wings than a Gaussian. This fits the observations well, better than
  the Gaussians in the top panel. The gas in the model has a large
  intrinsic dispersion, so the predictions do not show a double-peaked
  profile, despite the presence of a $3.3 \times 10^8 \Msun$
  BH.\label{f:lineshapes}}

\newcommand{\figcaprotfits}{Datapoints show the mean velocities $V$ (in $\kms$)
  inferred from Gaussian fits to the {\halnii} emission lines in the
  HST/FOS spectra. The solid curve shows the predictions for these
  data, for the model that provides the overall best fit to all
  available kinematical data (Figure~\ref{f:modelfits}). This model
  has $\Mbh = 3.3 \times 10^8 \Msun$. The long and short dashed curves
  shows the predictions of models that best fit the data when the BH
  mass is kept fixed at $\Mbh = 0$ and $\Mbh = 1.0 \times 10^9 \Msun$,
  respectively, yielding rotation curves that are either too shallow
  or too steep. Both models are ruled out at $>99$\% confidence
  (cf.~Figure~\ref{f:chisqv}). The predictions in this figure use our
  standard model for the stellar luminosity density ($\alpha = 0.5$).
  Similar results are obtained for other values of
  $\alpha$.\label{f:rotfits}}

\newcommand{\figcapchisqv}{Models were constructed with the BH mass fixed to
  the value shown along the abscissa. The remaining parameters in each
  model were chosen to optimize the fit to all available gas
  kinematical data (shown in Figure~\ref{f:modelfits}). The solid
  curve shows for each model thus obtained the quantity $\chi^2_V$
  that measures the quality of the fit to the HST/FOS rotation
  velocity measurements (shown in Figure~\ref{f:rotfits}). This
  quantity follows a $\chi^2$ probability distribution with $N_{\rm
  df} = 5$. Horizontal dashed lines at given values of $\chi^2_V$ show
  the probability that a value equal to or smaller than indicated
  would be observed for a correct model. The model without a BH is
  ruled out at $>99$\% confidence. The predictions in this figure use
  our standard model for the stellar luminosity density ($\alpha =
  0.5$).  Similar results are obtained for other values of $\alpha$
  (cf.~Figure~\ref{f:alphadep}).\label{f:chisqv}}

\newcommand{\figcapalphadep}{BH mass $\Mbh$ (in $\Msun$) as function of the
  central cusp slope $\alpha$ of the three-dimensional luminosity
  density. The heavy solid curve shows the $\Mbh$ that best fit the
  gas kinematics. The thin dotted, dashed, and long-dashed curves show
  the 68.3\%, 90\% and 99\% confidence intervals on $\Mbh$ implied by
  the gas kinematics, respectively. The heavy long-dashed curve shows
  the relation between $\Mbh$ and $\alpha$ implied by the assumption
  that the presence of a density cusp is due to adiabatic growth of a
  BH into a pre-existing homogeneous core. The horizontal arrows
  indicate the upper limits on $\alpha$ obtained from the observed
  {\BB}$-${\II} color, and from the fact that a galaxy with the
  luminosity of NGC 7052 is probably a core galaxy. Our standard model
  has $\alpha=0.5$ and $\Mbh = 3.3 \times 10^8 \Msun$. Models without
  a BH are ruled out with $>99$\% confidence for all values of
  $\alpha$ plotted here.\label{f:alphadep}}

\newcommand{\figcapallBHs}{Measurements of BH mass $\Mbh$ versus {\BB}-band
  spheroid luminosity $L_{B,{\rm sph}}$, for `secure' BH detections
  inferred from: ($\circ$)~ionized gas kinematics of nuclear disks;
  ($\bullet$)~stellar kinematical studies; ($\times$)~radio
  observations of water masers; ($\ast$)~time variability of broad
  double-peaked Balmer lines. The $\Mbh$ values are typically believed
  to be accurate to $|\Delta \log \Mbh | \lesssim 0.3$. The dashed
  line shows the $\Mbh$ for which the BH sphere of influence,
  $r_{\bullet} \simeq G \Mbh / \sigma^2$, extends $0.1''$ at a
  distance of $10 \Mpc$ ($\sigma$ is determined by $L_{B,{\rm sph}}$
  through the Faber-Jackson relation). BHs below this line can be
  detected only in galaxies closer than $10 \Mpc$, and in galaxies in
  which kinematical tracers can be observed at resolutions $< 0.1''$
  (e.g., water masers). This figure was adapted from figure~1a of van
  der Marel~(1998), which also lists the references for the individual
  galaxies.\label{f:allBHs}}


\ifsubmode
\figcaption{\figcapimages}
\figcaption{\figcapellipsefit}
\figcaption{\figcapminorprof}
\figcaption{\figcapminorcolor}
\figcaption{\figcapreddening}
\figcaption{\figcapcentralslope}
\figcaption{\figcapgasfit}
\figcaption{\figcapaperpos}
\figcaption{\figcapemlines}
\figcaption{\figcapgaskin}
\figcaption{\figcapmodelfits}
\figcaption{\figcaplineshapes}
\figcaption{\figcaprotfits}
\figcaption{\figcapchisqv}
\figcaption{\figcapalphadep}
\figcaption{\figcapallBHs}
\clearpage
\else\printfigtrue\fi

\ifprintfig


\clearpage
\begin{figure}
\epsfxsize=16.0truecm
\centerline{\epsfbox{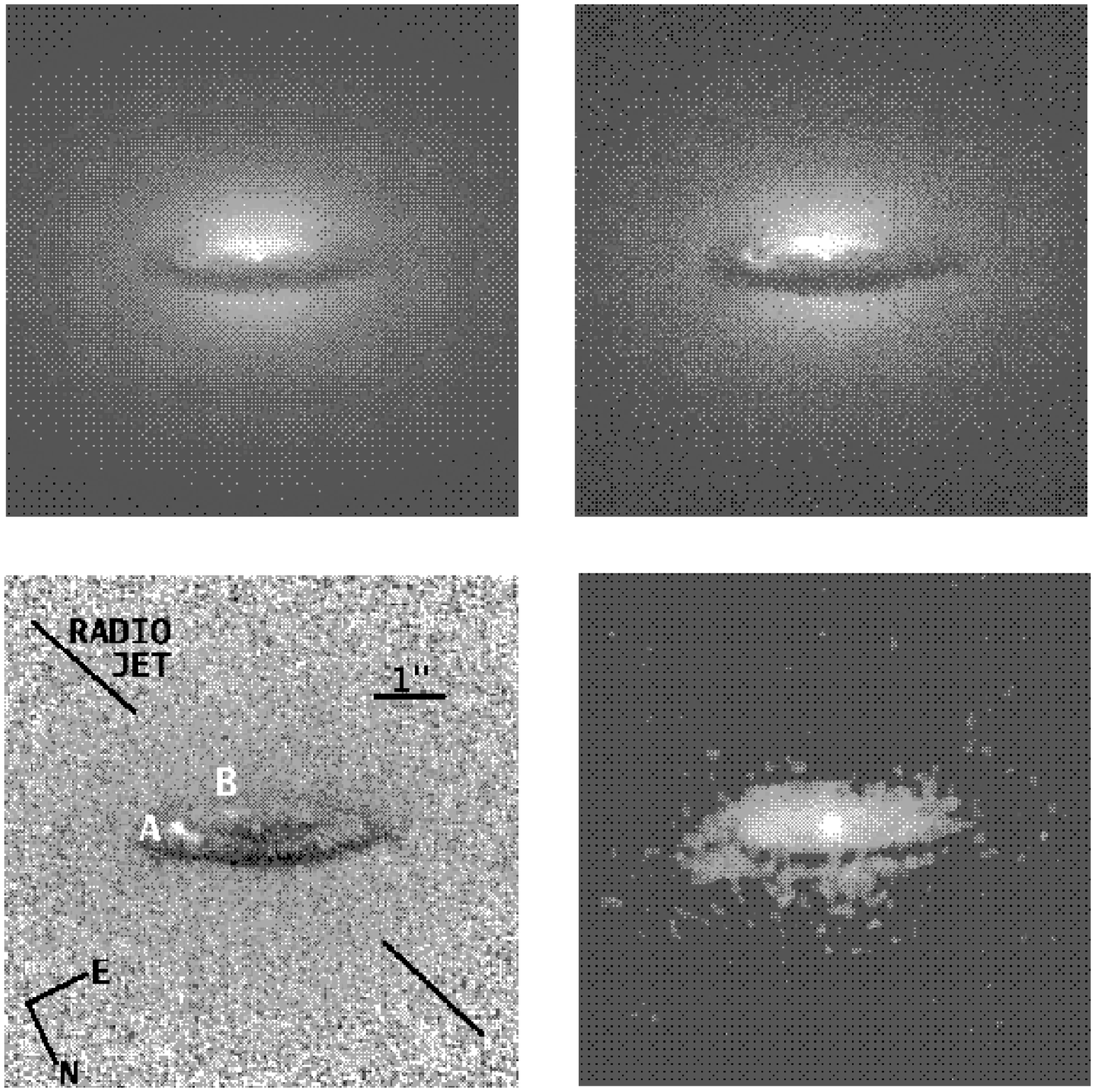}}
\ifsubmode
\vskip3.0truecm
\setcounter{figure}{0}
\addtocounter{figure}{1}
\centerline{Figure~\thefigure}
\else\vskip-0.3truecm\figcaption{\figcapimages}\fi
\end{figure}


\clearpage
\begin{figure}
\epsfxsize=16.0truecm
\centerline{\epsfbox{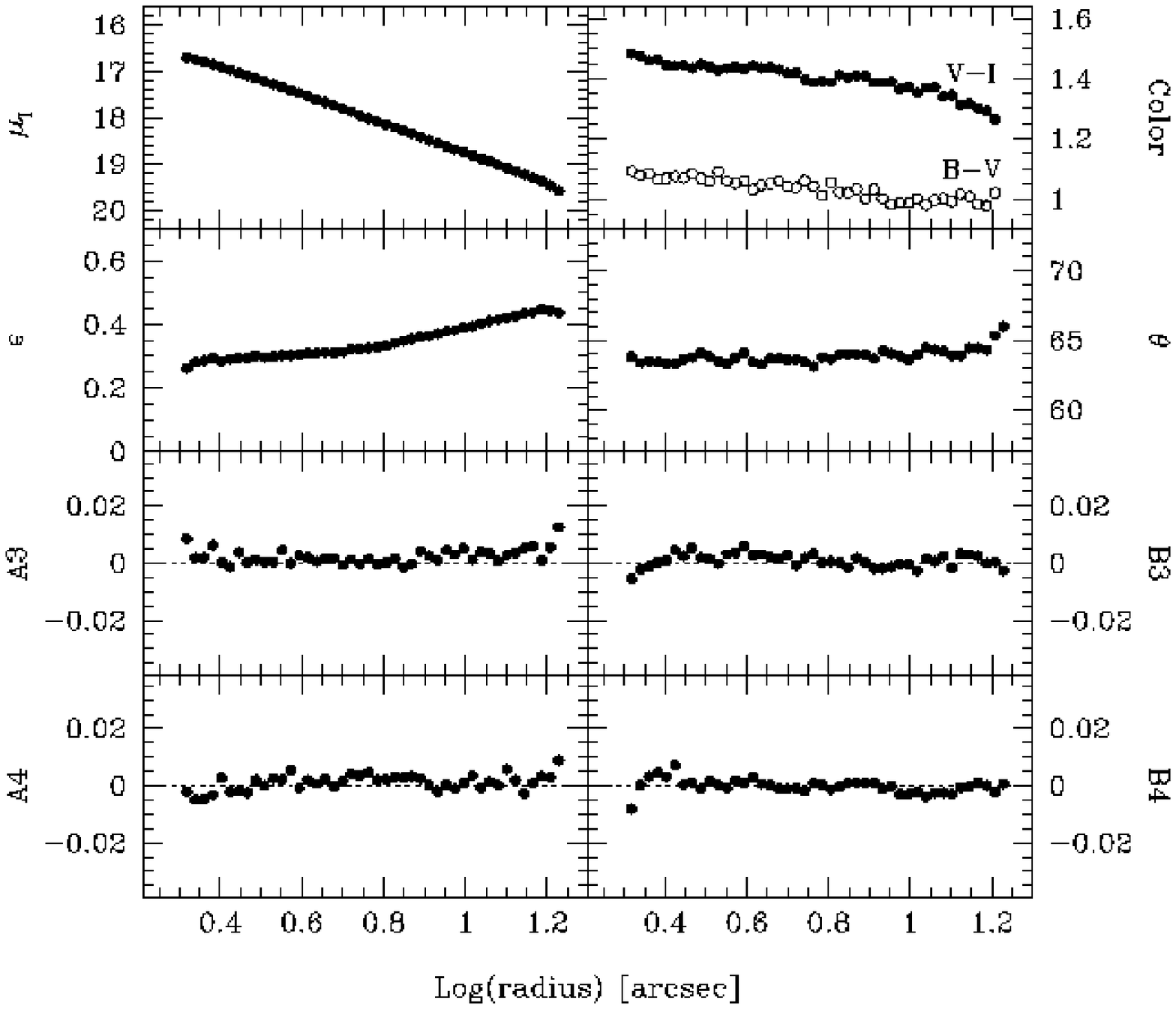}}
\ifsubmode
\vskip3.0truecm
\addtocounter{figure}{1}
\centerline{Figure~\thefigure}
\else\figcaption{\figcapellipsefit}\fi
\end{figure}


\clearpage
\begin{figure}
\epsfysize=14.0truecm
\centerline{\epsfbox{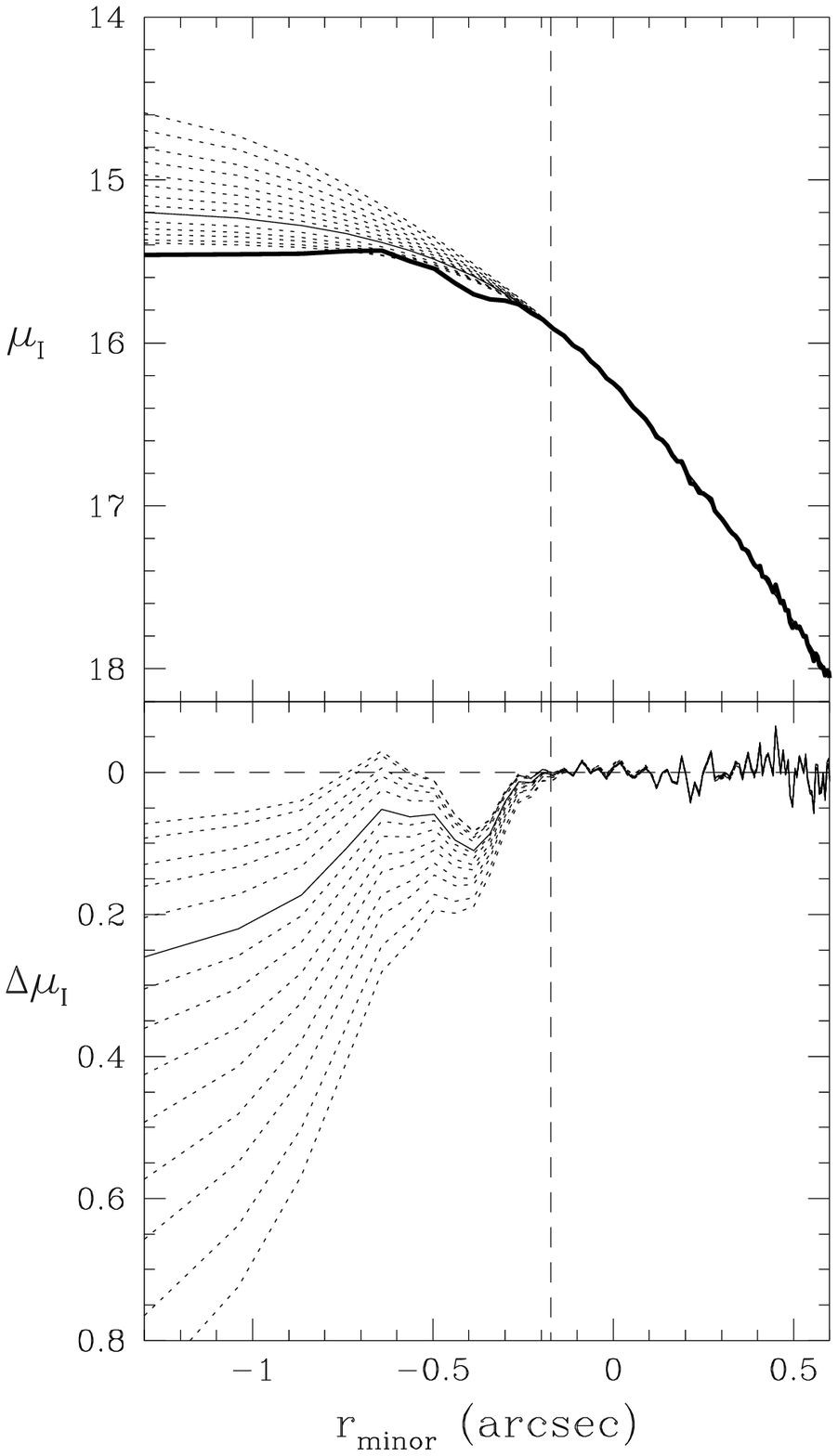}}
\ifsubmode
\vskip3.0truecm
\addtocounter{figure}{1}
\centerline{Figure~\thefigure}
\else\figcaption{\figcapminorprof}\fi
\end{figure}


\clearpage
\begin{figure}
\centerline{\epsfbox{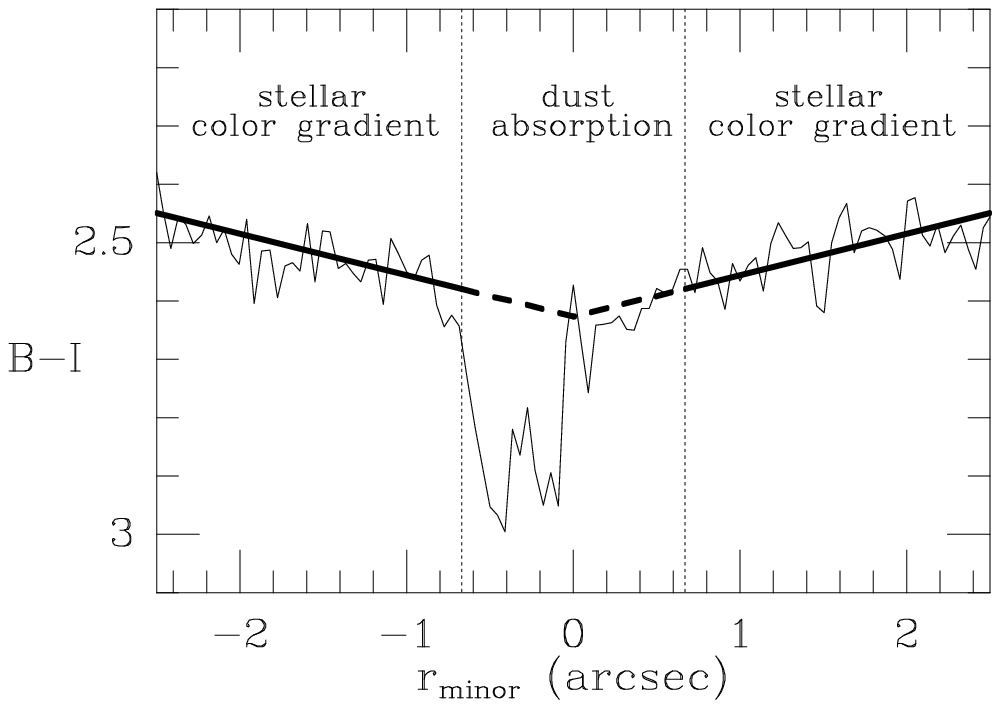}}
\ifsubmode
\vskip3.0truecm
\addtocounter{figure}{1}
\centerline{Figure~\thefigure}
\else\figcaption{\figcapminorcolor}\fi
\end{figure}


\clearpage
\clearpage
\begin{figure}
\epsfxsize=16.0truecm
\centerline{\epsfbox{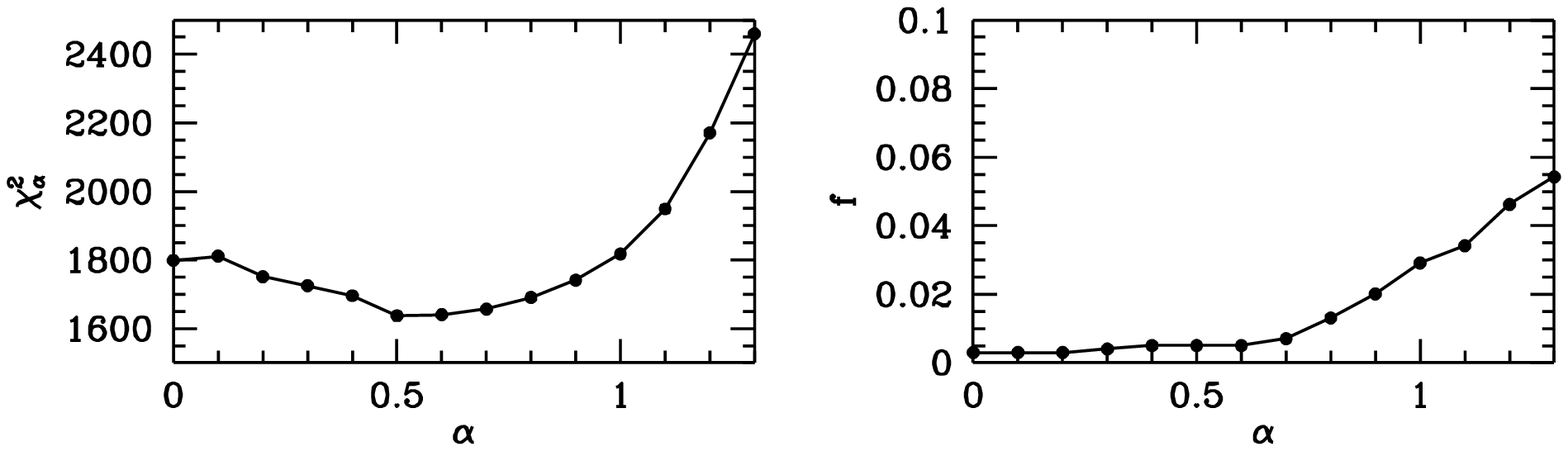}}
\ifsubmode
\vskip3.0truecm
\addtocounter{figure}{1}
\centerline{Figure~\thefigure}
\else\figcaption{\figcapreddening}\fi
\end{figure}


\clearpage
\begin{figure}
\centerline{\epsfbox{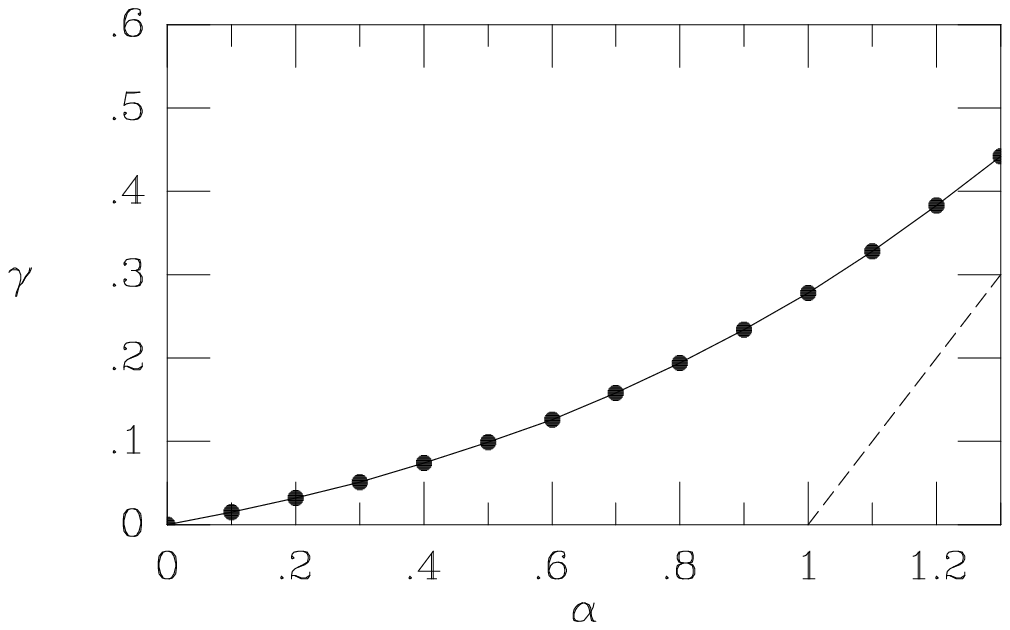}}
\ifsubmode
\vskip3.0truecm
\addtocounter{figure}{1}
\centerline{Figure~\thefigure}
\else\figcaption{\figcapcentralslope}\fi
\end{figure}


\clearpage
\begin{figure}
\epsfxsize=16.0truecm
\centerline{\epsfbox{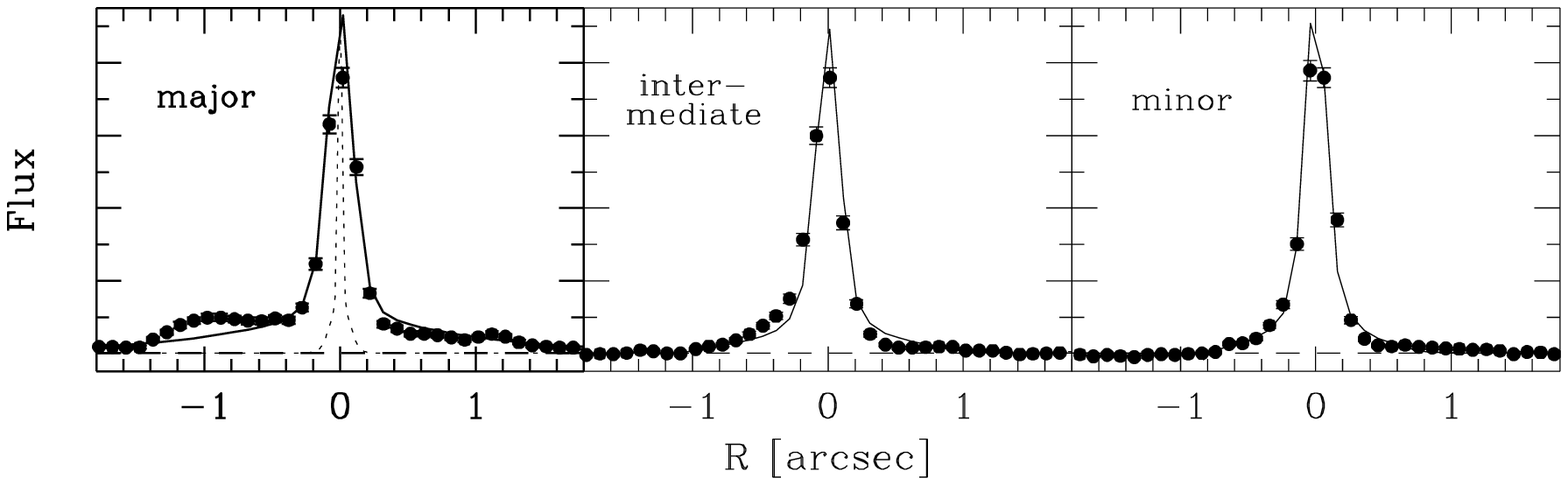}}
\ifsubmode
\vskip3.0truecm
\addtocounter{figure}{1}
\centerline{Figure~\thefigure}
\else\figcaption{\figcapgasfit}\fi
\end{figure}


\clearpage
\begin{figure}
\centerline{\epsfbox{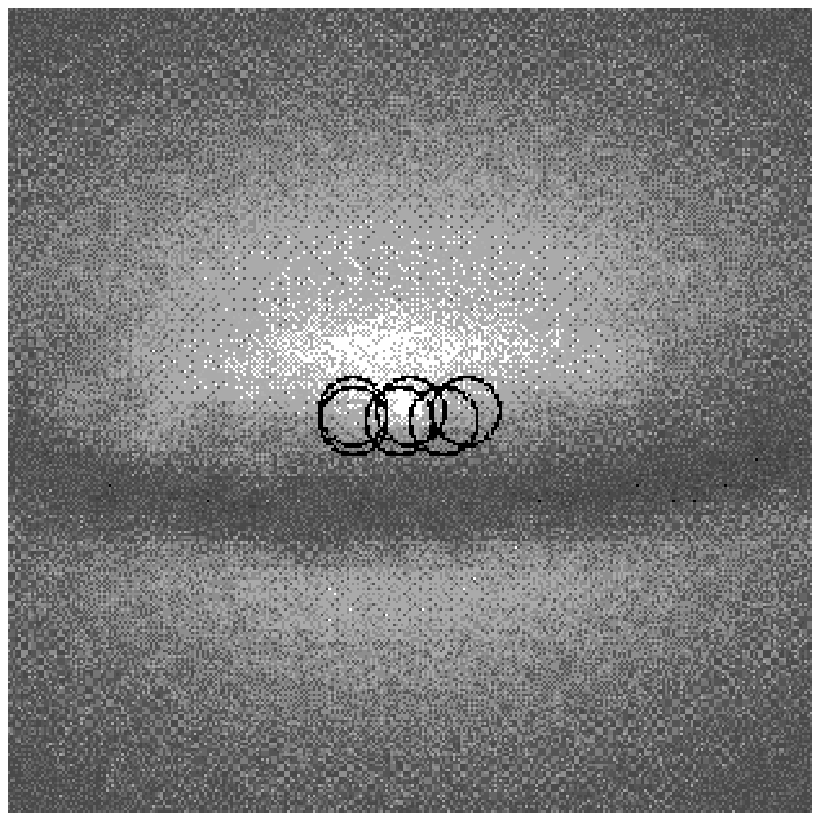}}
\ifsubmode
\vskip3.0truecm
\addtocounter{figure}{1}
\centerline{Figure~\thefigure}
\else\figcaption{\figcapaperpos}\fi
\end{figure}


\clearpage
\begin{figure}
\epsfxsize=16.0truecm
\centerline{\epsfbox{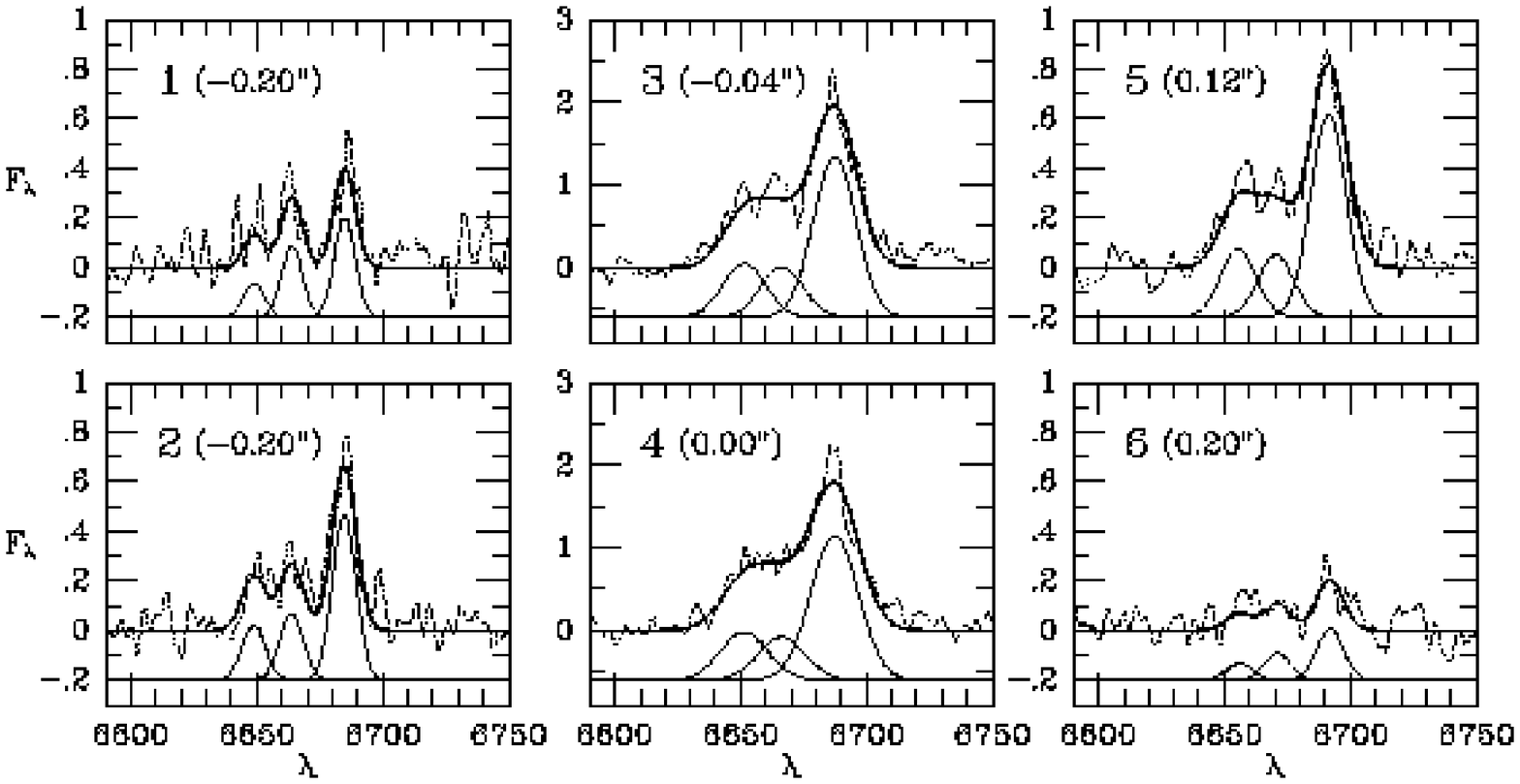}}
\ifsubmode
\vskip3.0truecm
\addtocounter{figure}{1}
\centerline{Figure~\thefigure}
\else\figcaption{\figcapemlines}\fi
\end{figure}


\clearpage
\begin{figure}
\centerline{\epsfbox{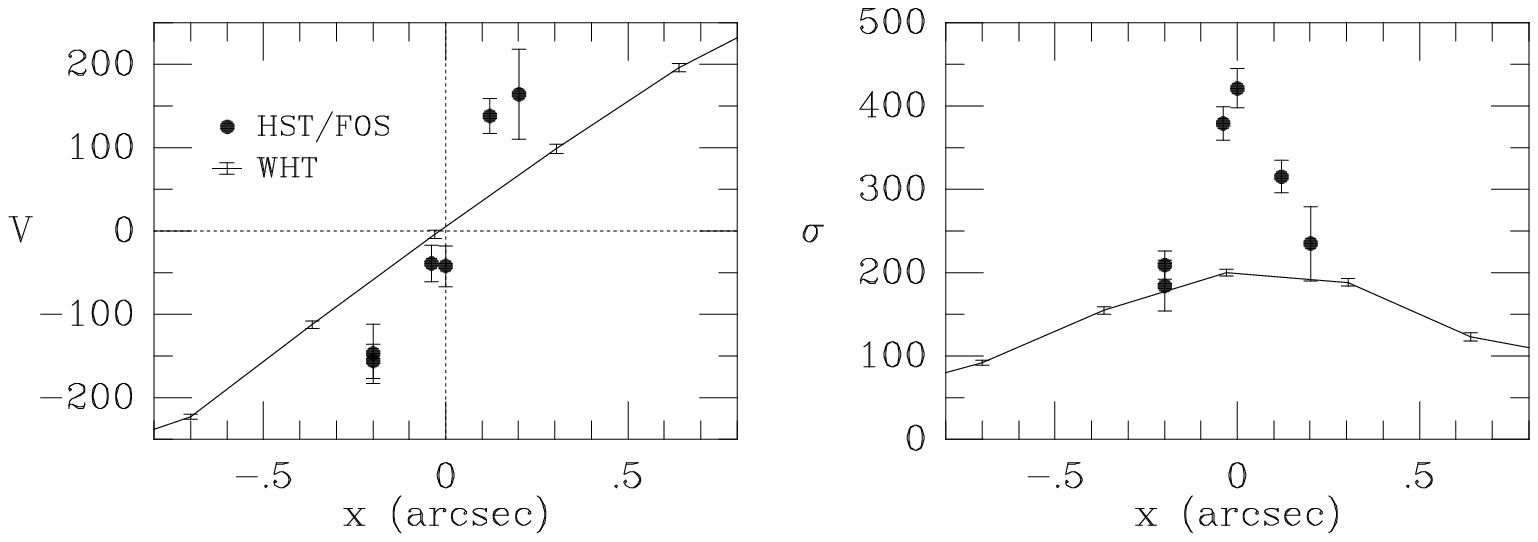}}
\ifsubmode
\vskip3.0truecm
\addtocounter{figure}{1}
\centerline{Figure~\thefigure}
\else\figcaption{\figcapgaskin}\fi
\end{figure}


\clearpage
\begin{figure}
\centerline{\epsfbox{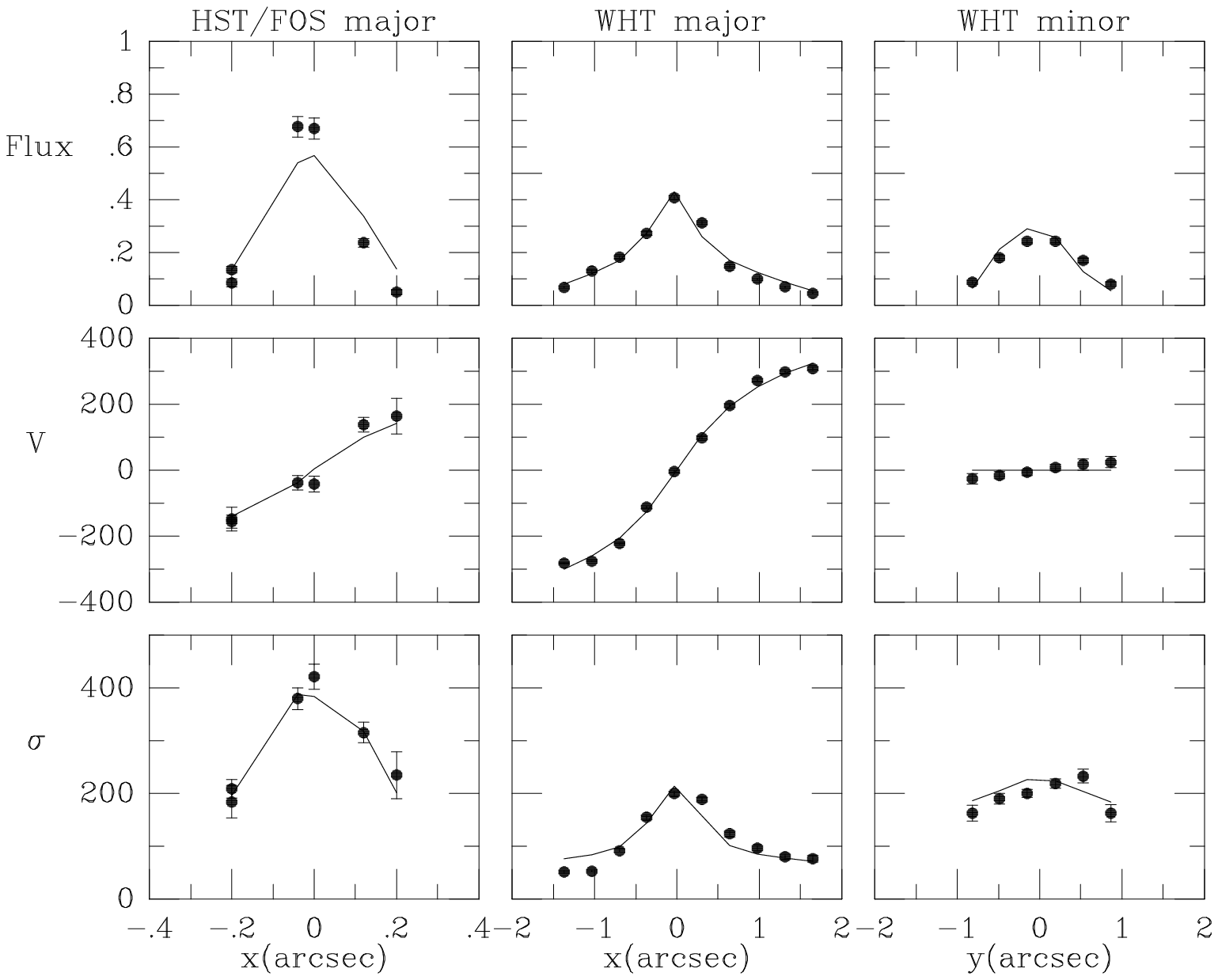}}
\ifsubmode
\vskip3.0truecm
\addtocounter{figure}{1}
\centerline{Figure~\thefigure}
\else\figcaption{\figcapmodelfits}\fi
\end{figure}


\clearpage
\begin{figure}
\centerline{\epsfbox{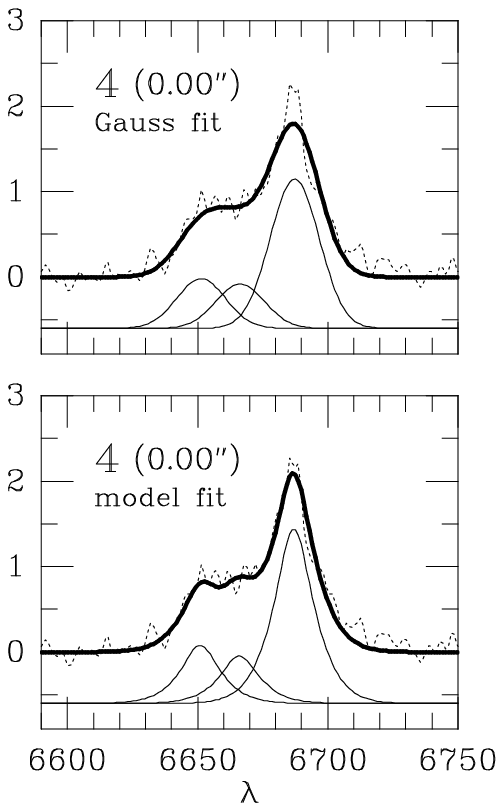}}
\ifsubmode
\vskip3.0truecm
\addtocounter{figure}{1}
\centerline{Figure~\thefigure}
\else\figcaption{\figcaplineshapes}\fi
\end{figure}


\clearpage
\begin{figure}
\centerline{\epsfbox{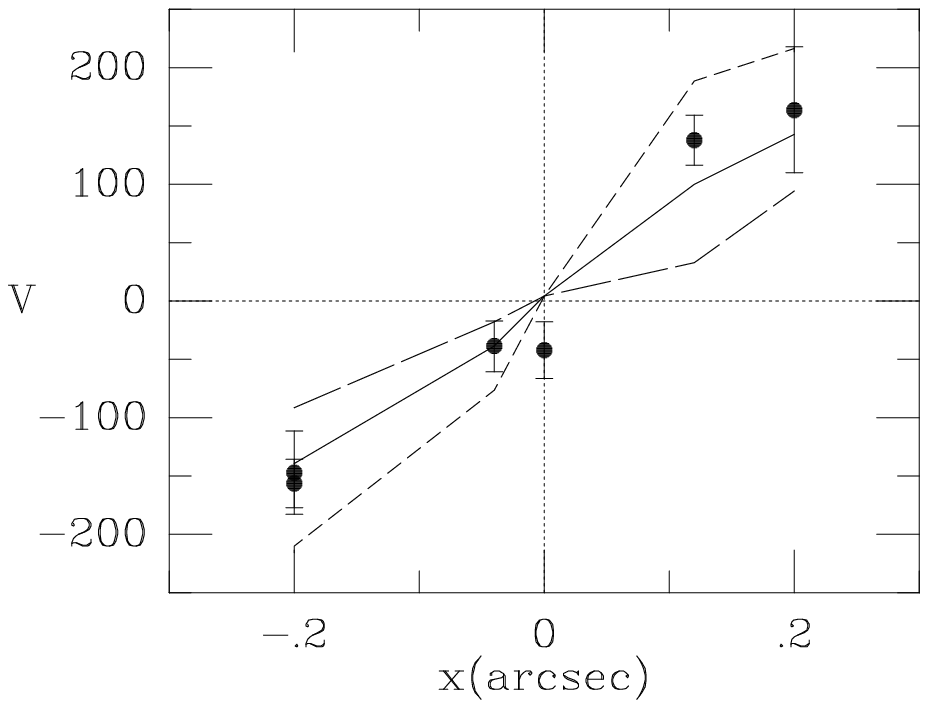}}
\ifsubmode
\vskip3.0truecm
\addtocounter{figure}{1}
\centerline{Figure~\thefigure}
\else\figcaption{\figcaprotfits}\fi
\end{figure}


\clearpage
\begin{figure}
\centerline{\epsfbox{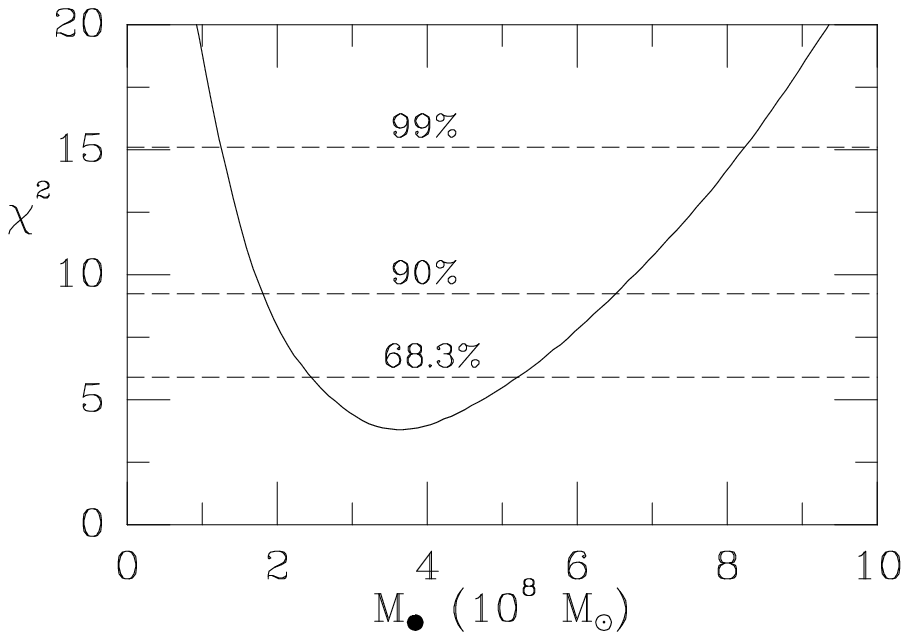}}
\ifsubmode
\vskip3.0truecm
\addtocounter{figure}{1}
\centerline{Figure~\thefigure}
\else\figcaption{\figcapchisqv}\fi
\end{figure}


\clearpage
\begin{figure}
\centerline{\epsfbox{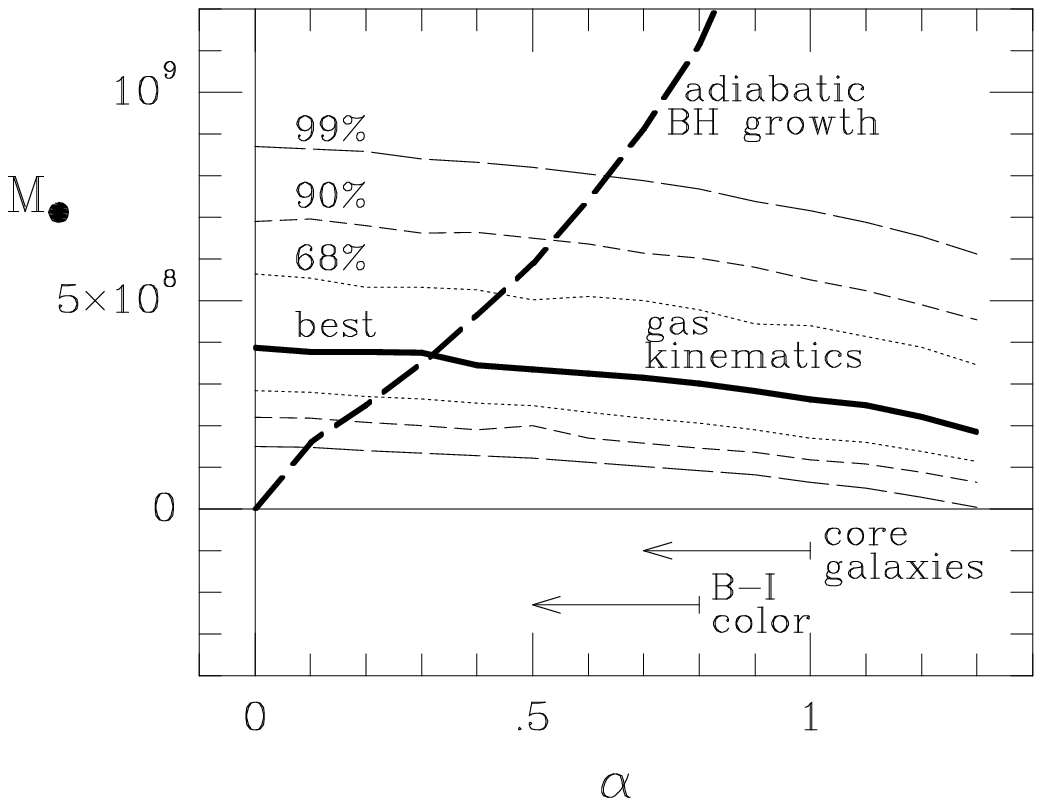}}
\ifsubmode
\vskip3.0truecm
\addtocounter{figure}{1}
\centerline{Figure~\thefigure}
\else\figcaption{\figcapalphadep}\fi
\end{figure}


\clearpage
\begin{figure}
\centerline{\epsfbox{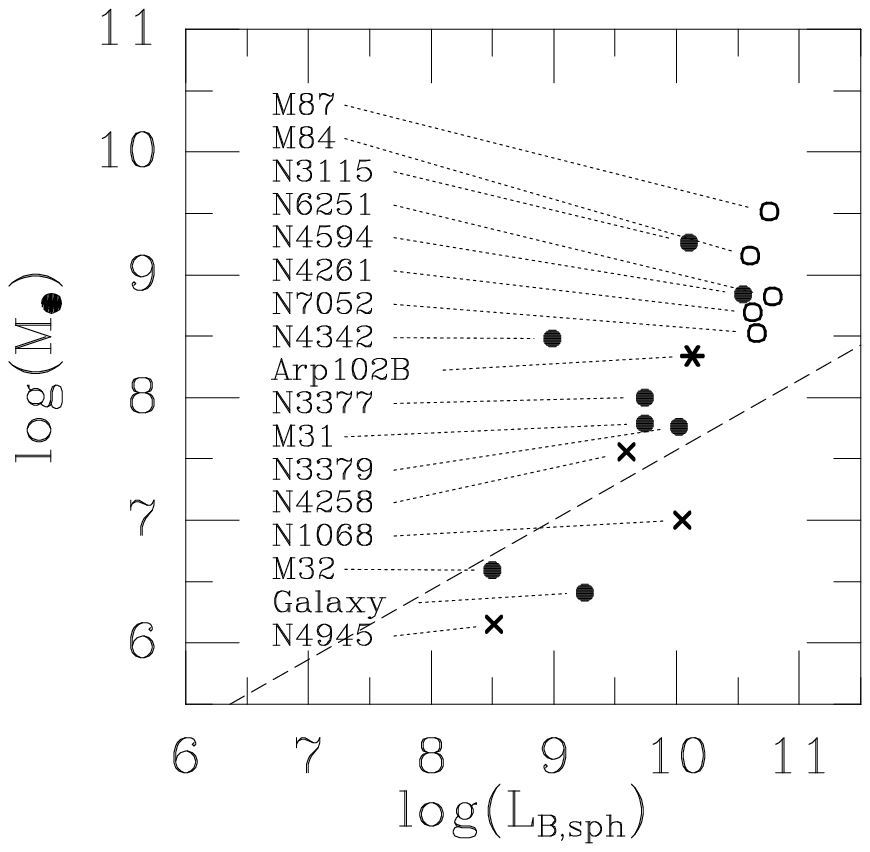}}
\ifsubmode
\vskip3.0truecm
\addtocounter{figure}{1}
\centerline{Figure~\thefigure}
\else\figcaption{\figcapallBHs}\fi
\end{figure}


\fi


\clearpage
\ifsubmode\pagestyle{empty}\fi


\begin{deluxetable}{lrrcrlc}
\tablecaption{HST/WFPC2 images: observational setup\label{t:images}}
\tablehead{
\colhead{filter} & \colhead{$\lambda_0$} & \colhead{$\Delta \lambda$} &
\colhead{HST-ID} & \colhead{$T_{\rm exp}$} & \colhead{CCD} & \colhead{pixel} \\
& \colhead {(\AA)} & \colhead{(\AA)} & & 
\colhead{(s)} & & \colhead{(arcsec)} \\
\colhead{(1)} & \colhead{(2)} & \colhead{(3)} & \colhead{(4)} &
\colhead{(5)} & \colhead{(6)} & \colhead{(7)} \\
}
\startdata
F450W          & 4445 &  925 & 4--6 & 1030 & PC  & $0.046$ \\
F547M          & 5454 &  487 & 7--9 &  840 & WF2 & $0.100$ \\
LRF(off-band)  & 6480 &   83 & d--f & 3600 & WF3 & $0.100$ \\
LRF(on-band)   & 6675 &   85 & a--c & 3700 & WF2 & $0.100$ \\
F814W          & 8269 & 1758 & 1--3 & 1470 & PC  & $0.046$ \\
\enddata
\tablecomments{WFPC2 images of NGC 7052 were obtained with 5 different 
filters. The filter name is listed in column~(1). LRF stands for
`linear ramp filter' (a filter with a tunable central
wavelength). Column~(2) and~(3) list the central wavelength of the
filter and the FWHM, as defined in Biretta \etal (1996). In the HST
Data Archive, the observations have names of the form
u2p9010$\ast$t. Column~(4) identifies the $\ast$ in the Archive name
for each observation. Column~(5) lists the total exposure time per
filter, which for each filter was divided over three different
exposures. Column~(6) lists the WFPC2 CCD on which the target was
positioned, and column~(7) lists the resulting pixel size.}
\end{deluxetable}


\begin{deluxetable}{cccrrrrrrrr}
\tablecaption{HST/FOS spectra: observational setup and gas 
kinematics\label{t:spectra}}
\tablehead{
\colhead{ID} & \colhead{date} & \colhead{HST-ID} &
\multicolumn{2}{c}{position} & \colhead{$T_{\rm exp}$} &
\colhead{$V$} & \colhead{$\Delta V$} & 
\colhead{$\sigma$} & \colhead{$\Delta \sigma$} \\ 
 & & & \colhead{$x$ (arcsec)} & \colhead{$y$ (arcsec)} & \colhead{($s$)} &
\colhead{km/s} & \colhead{km/s} & \colhead{km/s} & \colhead{km/s} \\
\colhead{(1)} & \colhead{(2)} & \colhead{(3)} & 
\colhead{(4)} & \colhead{(5)} & \colhead{(6)} &
\colhead{(7)} & \colhead{(8)} & \colhead{(9)} & \colhead{(10)} \\
}
\startdata
1 & Aug 96 & b       & $-0.20$ & $-0.02$ & 2380 & $-147$ & 36 & 184 & 31 \\
2 & Sep 95 & d, f    & $-0.20$ & $0.01$  & 3830 & $-156$ & 21 & 209 & 17 \\
3 & Aug 96 & d       & $-0.04$ & $-0.02$ & 2380 & $-39$  & 22 & 379 & 20 \\
4 & Sep 95 & 7       &  $0.00$ & $0.01$  & 2370 & $-42$  & 24 & 421 & 23 \\
5 & Aug 96 & 7, 9, f &  $0.12$ & $-0.02$ & 6220 & $138$  & 21 & 315 & 19 \\
6 & Sep 95 & 9, b    &  $0.20$ & $0.01$  & 4770 & $164$  & 54 & 235 & 44 \\
\enddata
\tablecomments{FOS spectra of NGC 7052 were obtained at six different 
aperture positions. Column~(1) is the label for the spectrum used in
the remainder of the paper. Column~(2) lists the month in which the
obervation was obtained. In the HST Data Archive, the observations in
September 1995 have names of the form y2p9020$\ast$p; those in August
1996 have names of the form y2p9040$\ast$t. Column~(3) identifies the
$\ast$ in the Archive name for each observation. Columns~(4) and~(5)
list the aperture position for each observation, determined as
described in Appendix~\ref{s:AppA}. The $(x,y)$ coordinate system is
centered on the galaxy, with the $x$-axis along the major axis
(position angle $63.5^{\circ}$). Column~(6) lists the exposure
time. Columns~(7)-(10) list the mean velocity $V$ and velocity
dispersion $\sigma$ of the emission line gas, with corresponding
formal errors, determined from Gaussian fits to the emission lines as
described in the text.}
\end{deluxetable}


\end{document}